\documentclass{article}
\usepackage{setspace}
\usepackage[margin=3cm,nohead]{geometry}
\usepackage{appendix,graphicx}
\usepackage{amsfonts,amssymb,amsmath,mathenv}
\allowdisplaybreaks[3]

\newcommand{\nc}{\newcommand}
\def\foot{\footnote}
\def \bi{\bibitem}
\def \ci{\cite}
\nc{\sr}{\sqrt}
\nc{\fr}{\frac}
\nc{\ov}{\over}
\nc{\x}{\times}
\nc{\cosec}{\textrm{\,cosec\,}} 
\nc{\sech}{\textrm{\,sech\,}}
\nc{\cosech}{\textrm{\,cosech\,}}
\nc{\del}{\partial}
\nc{\dpl}{\partial_+}
\nc{\dm}{\partial_-}
\nc{\dpp}{\partial_+\partial_+}
\nc{\dmm}{\partial_-\partial_-}
\nc{\dpm}{\partial_+\partial_-}
\nc{\ra}{\rightarrow}
\nc{\lra}{\leftrightarrow}
\nc{\Ra}{\Rightarrow}
\nc{\til}{\tilde}
\nc{\R}{\mathbb{R}}
\nc{\Z}{\mathbb{Z}}
\nc{\C}{\mathbb{C}}
\nc{\al}{\alpha}
\nc{\bet}{\beta}
\nc{\ga}{\gamma}
\nc{\de}{\delta}
\nc{\h}{\eta}
\nc{\thet}{\theta}
\nc{\ka}{\kappa}
\nc{\lam}{\lambda}
\nc{\si}{\sigma}
\nc{\Si}{\Sigma}
\nc{\ta}{\tau}
\nc{\ze}{\zeta}
\nc{\y}{\psi}
\nc{\om}{\omega}
\nc{\vt}{\vartheta}
\nc{\vp}{\varphi}
\nc{\be}{\begin{equation}}
\nc{\ee}{\end{equation}}
\nc{\bd}{\begin{displaymath}}
\nc{\ed}{\end{displaymath}}
\nc{\ba}{\begin{array}}
\nc{\ea}{\end{array}}
\nc{\la}{\label}
\def \rf{\eqref}
\nc{\nin}{\noindent}
\nc{\hs}{\hspace}
\nc{\vs}{\vspace}
\nc{\np}{\vspace{10pt} \\}
\nc{\mc}{\mathcal}
\nc{\mf}{\mathfrak}
\nc{\mbb}{\mathbb}
\nc{\trm}{\textrm}
\nc{\tbf}{\textbf}
\nc{\mbf}{\mathbf}
\def \Str {{\rm STr}}
\nc{\Lag}{\mathcal{L}}
\nc{\Ham}{\mathcal{H}}
\nc{\psu}{\mf{psu}\left(2,\,2\,|\,4\right)}
\nc{\Psu}{PSU\left(2,\,2\,|\,4\right)}
\nc{\gi}{g^{-1}}
\nc{\finv}{f^{-1}}
\nc{\YL}{\Psi_{_L}}
\nc{\YR}{\Psi_{_R}}
\nc{\YLo}{\Psi_{_L0}}
\nc{\YRo}{\Psi_{_R0}}
\nc{\hm}{\h^\parallel}
\nc{\hh}{\h^\perp}
\nc{\kasym}{$\ka$-symmetry }
\nc{\T}{\mc{T}}
\newcommand{\commut}[2]{\left[ #1{,}\,#2 \right] }

\def \ha {\fr{1}{2}}
\def \ept {e^{i{\fr{\ka^2- m^2}{\ka} \ta}}}

\def \la {\label} 
\def \tvt {\vartheta'}
\def \adss {$AdS_5 \times S^5$ } 
\def \bea {\begin{eqnarray}}
\def \eea {\end{eqnarray}}
\def \no {\nonumber}
\def \s {\sigma} \def \p {\phi}
\def \m {\mu} \def \k {\kappa}
\def \const {{\rm const}}

 \def \mm {\ell}

\begin{document}

\overfullrule=0pt
\parskip=2pt
\parindent=12pt
\headheight=0in \headsep=0in \topmargin=0in \oddsidemargin=0in

\vspace{ -3cm}
\thispagestyle{empty}
\vspace{-1cm}

\rightline{Imperial-TP-AT-2009-2}

\rightline{    }

\begin{center}
\vspace{1cm}
{\Large\bf  
Pohlmeyer-reduced  form of   string theory  in $AdS_5\times S^5$: \\ 
semiclassical expansion 


\vspace{1.2cm}


   }

\vspace{.2cm} {
B. Hoare\footnote{ benjamin.hoare08@imperial.ac.uk }, 
Y. Iwashita\footnote{ yukinori.iwashita07@imperial.ac.uk }  and A.A. Tseytlin\footnote{Also at Lebedev  Institute, Moscow. tseytlin@imperial.ac.uk }}\\

\vskip 0.6cm

{\em 
 Theoretical Physics Group \\
 Blackett Laboratory, Imperial College\\
London SW7 2AZ, U.K. }

\end{center}

\onehalfspacing
\setcounter{footnote}{0}
\begin{abstract}
 We consider the Pohlmeyer-reduced formulation of the
 \adss  superstring. It is constructed by 
 introducing new variables  which  are algebraically related to 
 supercoset current components so that the Virasoro 
 conditions are automatically solved.  The  reduced theory 
 is  a gauged  WZW  model  supplemented with an integrable potential  and  fermionic terms 
 that ensure its UV finiteness.  The  original  superstring theory 
 and  its reduced counterpart   are  closely related  at the  classical level, 
 and we conjecture 
 that they remain related at  the quantum level as well, 
  in the sense that their quantum partition functions evaluated on respective 
 classical solutions  are equal. 
 We provide evidence for  the validity  of this 
 conjecture at the  one-loop level, i.e. at the first non-trivial 
 order of the semiclassical  expansion near several classes  of classical solutions.

\end{abstract}

\def \adss {$AdS_5 \times S^5$ \ } 

\newpage
\tableofcontents

\setcounter{footnote}{0}

\renewcommand{\theequation}{1.\arabic{equation}}
\setcounter{equation}{0}
 \newpage
\section{Introduction\la{Intro}}

In this paper  we continue the investigation of the Pohlmeyer-reduced   form 
of the \adss  superstring theory  initiated in \ci{gt1,mikhshaf,gt2,rtfin}.

The original  Pohlmeyer reduction procedure, \ci{pol}, relates the  classical equations of motion 
of the 
sigma model  on $S^2$ to the  sine-Gordon equation. The reduction may be interpreted, \ci{tse,mi1,mi2},
as  solving the Virasoro conditions in the classical conformal-gauge  string theory 
on $\mbb{R}_t \times S^2$ (with the residual conformal diffeomorphisms fixed by $t= \mu \tau$ condition) 
in terms  of the remaining  physical   degree of freedom: identified  as the angle variable of the 
sine-Gordon  model. This relation between the $S^2$ sigma model  and  the sine-Gordon model 
(and its generalizations to other  similar bosonic  sigma models)
was  used  for  explicit construction of  several interesting   classical   string solutions
on symmetric spaces like $S^n$  and $AdS_n$ 
(see, e.g.,  \ci{deve,hm,dor,okamura,hosv,finsizemag2,jevi,mira,dorn,ald}).

An attractive feature of the Pohlmeyer-reduced  form of the  string theory  sigma model 
is that while  it involves only the physical (``transverse'')  degrees of freedom 
it  still has  manifest  2-d Lorentz  invariance.
It would  be very  useful to have such a  formulation  
for the  quantum \adss  string theory. 

Starting  with the equations of motion 
of the \adss superstring  described by the
 $\fr{F}{G} = \fr{\Psu}{Sp\left(2,\,2\right)\x Sp\left(4\right)}$ 
supercoset, which may be written 
 in terms of the 
$\Psu$ current one may  solve the Virasoro conditions  by introducing the new variables 
$g \in G=  Sp(2,2) \times Sp(4), \ A_\pm \in \mf{h}$,\foot{$\mf{h}$ is the Lie algebra of 
the subgroup  $H=SO(4) \times SO(4)$=$[SU(2)]^4$ of the group $ G$.}
 and $\Psi_{_{L,R}}$, 
which are algebraically related to the current components. 
The resulting equations   can then be  obtained 
from a local action $I_{\rm red}(g, A_\pm, \Psi_{_{L,R}})$
which happens to be the $G/H$ gauged WZW   model  modified by an $H$-invariant potential 
and supplemented  by the 2-d fermionic terms  (see \ci{gt1} and \rf{Ltot},\rf{gauwzw}  below). 
This action, which defines the reduced theory, is 2-d Lorentz invariant and 
(after  fixing the residual $H$ gauge symmetry) involves only the physical number (8+8)
of bosonic and fermionic degrees of freedom. 

The original \adss  superstring  theory and the reduced theory 
are essentially equivalent   at the classical level, having closely related integrable structures and 
sets of classical  solutions. The question that we would like to address here is 
 if this 
 correspondence may extend  to the quantum level.  
 
Since the classical  Pohlmeyer reduction utilizes 
conformal invariance, it has a chance to apply at the quantum level only 
if the sigma model one starts  with  is UV finite. This is the  case  for the 
 \adss  superstring sigma model, \ci{mt,rtt,rt2,review},  which 
 is a combination of the $AdS_5$  and the $S^5$ sigma models  ``glued'' together by  the Green-Schwarz fermions
 into a conformal 2-d theory.
 For consistency, the  corresponding  reduced theory,  \ci{gt1,mikhshaf}, 
 should   also be  UV finite. That was indeed 
 shown to be  true  to the two loop orders  and is
 expected to be true also  to all  orders, \ci{rtfin}.

It should be emphasized that we  are interested in the reduced theory 
only as a tool for describing  observables of the original string theory:
it is the string theory that should  dictate those quantities one should   compute 
in   the reduced theory.\foot{In particular, one  may  not  be able  to translate
some characteristics  of solitons in the reduced theory directly 
 into meaningful quantities in string theory, etc.
For example, the energies  of the corresponding 
solutions  in the reduced theory and in the string theory may be related (if at all) 
in a nontrivial way (cf. \ci{hm}).}
  
Since the construction  of the reduced theory from string theory  equations 
of motion involves  rewriting the theory in terms of the currents, the original superstring 
coordinates are effectively  non-local functions of  the new  reduced theory variables. 
As was noticed  in \ci{gt1}, the part of the reduced theory action 
given by the sum of the  bosonic  interaction potential and  the fermionic ``Yukawa'' term  is  
essentially the same  as  the original \adss  GS action expressed in terms of the new variables. 
This  suggests   that the two  theories may actually be related  by  a non-trivial change of 
variables (from fields to currents) 
 in the path integral, similar to the one  used   in  the  non-abelian duality transformations
(cf. \ci{ft,mi2}).

More precisely,    the  string theory  path integral  should contain delta-functions 
of the Virasoro constraints,  $\delta( T_{++}) \delta( T_{--})$, 
  and the  change of variables  from 
the supercoset coordinates to  currents  and to the reduced theory  fields 
 should solve  these constraints. 
 Heuristically,  the additional gauged WZW  and ``free'' fermionic terms
 present  in the reduced theory action may 
originate from the  functional Jacobian  of this change  of variables.  

\

With this motivation in mind, here we   propose the conjecture that  the 
quantum string theory partition function 
(e.g., on a plane  or on a cylinder)  should be equal  to the 
quantum reduced theory partition function,
\be 
\mc{Z}^{(q)}_{\bf string \ theory} = \mc{Z}^{(q)}_{\bf reduced\ theory}   \la{eq} \ .  \ee  
Since  these two theories have the same number (8+8) of  independent degrees of freedom 
this equality is obviously true in the trivial vacuum (BMN) case.

The aim of this   paper is to provide  evidence  for this conjecture 
in the one-loop approximation, i.e.  by expanding both sides  of \rf{eq}
near the corresponding classical solutions  and computing the 
determinants of the  quadratic fluctuation operators.\foot{The
classical  parts  of the partition  functions   determined by  the values of the  actions 
evaluated on the respective solutions  will  not match in general.
 The values of the two classical actions do not  coincide on generic solutions
 which may not be surprising  if part of the reduced theory action may be  
 indeed interpreted as coming from the
 Jacobian of change of variables in the path integral.
 This is  not a problem as the  value of the reduced theory action
 on a classical solution 
 is  not necessarily an observable that one  may  be  
  interested in on the string theory side.}

Given the classical equivalence between  the string theory and the reduced theory 
the relation  between the  one-loop corrections which are 
determined  by the quadratic 
fluctuation spectra  may not look too surprising:  after all, the quadratic fluctuation 
operators  can be found from the classical equations of motion
and thus should be expected to  be in correspondence.  However, 
given that the reduction procedure involves  nontrivial steps of  
non-local change of variables and partial gauge fixing  the  general  proof of the equivalence of the
one-loop  partition functions  defined  directly  by  the two actions  appears 
to be non-trivial (and will not be attempted here).  

Below we shall explicitly verify   \rf{eq} in the one-loop  approximation  for a 
 few  simple  classes of string solutions and their counterparts in 
the reduced theory:  (i)  generic string solutions localized in $AdS_2 \times S^2$ part of $AdS_5 \times S^5$, 
and (ii) the homogeneous string solution representing a spinning string  in $S^3$ part of $S^5$. 


\

 We shall start in  section 2  with a review 
 of the classical Pohlmeyer reduction 
 for the $AdS_5 \x S^5$   superstring theory 
  following  \ci{gt1}.  We shall mention the possibility 
  of  introducing  an automorphism $\tau$ of the algebra of $H$  in the construction  of the reduced 
  theory action (which then  generalizes to an  asymmetrically gauged $G/H$ WZW model) 
  and also comment on the vacuum structure of the reduced theory, (section 2.2). 


 In section 3 
 we will  consider the quadratic  fluctuations of the
  conformal-gauge string theory equations of  motion 
around  classical string solutions.  Here the fluctuating fields  are  string coordinates
 rather than currents but one can parametrize the dependence on the 
 classical background 
   in terms of the classical values of the  current components.
  This  allows  one to start  with a  classical solution  of the reduced theory and find 
  the  string  fluctuation equations near the corresponding classical string solution. 
We shall apply this procedure 
 to the case of  generic  $AdS_2 \times S^2$ string solutions, (section 3.2), 
preparing the ground for comparing with the  fluctuation  spectrum  in the reduced theory.

In section 4 we will   start with the action of the reduced  theory and expand
it to quadratic order near its  classical solution.  We  will  then specialize 
to the   case of the  reduced theory background corresponding to the  generic 
string theory solution localized in $AdS_2 \times S^2$
 subspace of $AdS_5 \x S^5$ (section 4.2).   Comparing to the quadratic fluctuation 
operators  found   on the string theory side in section 3 we will 
then  be able to  conclude that they match and thus \rf{eq}
  should be true at least in the one-loop approximation. 
The same conclusion will be reached in the case of 
 homogeneous string solutions  in $\mbb{R}_t \times S^3$ and in 
 $AdS_3 \times S^1$   (section 4.3).

Section 5  will contain a summary and  remarks on  open  problems.

In appendix A we will summarise some definitions and notation 
related to    $\Psu$  supergroup  and discuss decompositions of the corresponding superalgebra. 
In appendix B we shall relate the parametrization of the supercoset
 $\fr{\Psu}{Sp\left(2,\,2\right)\x Sp\left(4\right)}$ 
to standard embedding coordinates in $AdS_5 \times S^5$.
In appendix C we shall discuss some special cases of  string solutions localized 
in $AdS_2 \times S^2$ part of $AdS_5 \times S^5$ and  the corresponding  fluctuation 
equations  in the reduced theory. 
Appendix D will contain a brief discussion of reduced theory counterparts of simple homogeneous 
string solutions. 
In appendix E we shall discuss an alternative way of computing 
the bosonic fluctuation frequencies in the reduced
 theory, using as an example the homogeneous solution discussed in section 4.3.1.


\renewcommand{\theequation}{2.\arabic{equation}}
 \setcounter{equation}{0}

\section{Review of  the Pohlmeyer reduction of the \adss superstring\la{Class}}

In this section we  shall  give a 
 a brief summary of the classical Pohlmeyer reduction 
 for Type IIB superstring theory on $AdS_5 \x S^5$ 
 that follows \ci{gt1}. 

 We start with the  2-d worldsheet sigma model arising from the 
Green-Schwarz action for the
Type IIB superstring theory on $AdS_5 \x S^5$ after fixing 
 the conformal gauge. 
This is the $F/G$ coset sigma model 
where $F=\Psu$ and $G=Sp\left(2,\,2\right)\x Sp\left(4\right)$
(see appendix \ref{PSU}); we  will 
henceforth  call this   sigma model 
the  conformal-gauge string theory. 

  Let us  consider the  field $f \in \Psu $
and define the left-invariant current $J=\finv d f$. Under the $\Z_4$ decomposition discussed 
in appendix \ref{PSU} the current can be written  as follows
\be
J=\finv d f=\mc{A}+Q_1+P+Q_2\,,\hs{40pt}\mc{A}\in\mf{g},\;Q_1\in\mf{f}_1,\;P\in\mf{p},\;Q_2\in\mf{f}_3\,.
\ee 
The GS action in the conformal gauge is then
\be\la{fulllag}
L_{GS}=\trm{STr}\Big[P_+P_-+\fr{1}{2}(Q_{1+}Q_{2-}-Q_{1-}Q_{2+})\Big]\,,
\ee
where $\partial_\pm=\partial_\tau\pm\partial_\sigma$. 
We also need to impose the conformal-gauge (Virasoro) constraints,
\be\la{vir}
\Str\left(P_\pm P_\pm\right)=0\,.
\ee
This system has a $G$ gauge symmetry under which,
\be\ba{c}
f\ra fg \hs{10pt} \Ra \hs{10pt} J\ra \gi J g+\gi d g\,,
\\\hs{130pt}\Ra \hs{10pt} P\ra \gi P g\,, \hs{20pt} \mc{A}\ra \gi \mc{A} g+\gi dg\,,
\\\hs{150pt} Q_1\ra \gi Q_1 g\,, \hs{20pt} Q_2\ra \gi Q_2 g\,.
\ea\ee

The equations of motion of the conformal-gauge string theory, obtained 
 by varying $f$ in \eqref{fulllag}, are
\be\ba{c}\la{fulleom}
\dpl P_- +\left[\mc{A}_+,P_-\right]+\left[Q_{2+},Q_{2-}\right] =0\,,
\\\dm P_+ +\left[\mc{A}_-,P_+\right]+\left[Q_{1-},Q_{1+}\right]=0\,,
\\\left[P_+,Q_{1-}\right]=0\,,\hs{20pt}\left[P_-,Q_{2+}\right]=0\,.
\ea\ee
Interpreted as equations for the current  components 
they should be supplemented by the Maurer-Cartan equation
\be\la{mc}
\dm J_+ -\dpl J_- +\left[J_-,J_+\right]=0\,.
\ee
Under the $\Z_4$ decomposition the Maurer-Cartan equation \rf{mc} takes the form 
\be\ba{c}\la{decompmc}
\dm P_+ -\dpl P_- +\left[\mc{A}_-,P_+\right]+\left[Q_{1-},Q_{1+}\right]+\left[P_-,\mc{A}_+\right]+\left[Q_{2-},Q_{2+}\right]=0\,,
\\\dm \mc{A}_+ -\dpl \mc{A}_- +\left[\mc{A}_-,\mc{A}_+\right]+\left[Q_{1-},Q_{2+}\right]+\left[P_-,P_+\right]+\left[Q_{2-},Q_{1+}\right]=0\,,
\\\dm Q_{1+} -\dpl Q_{1-} +\left[\mc{A}_-,Q_{1+}\right]+\left[Q_{1-},\mc{A}_+\right]+\left[P_-,Q_{2+}\right]+\left[Q_{2-},P_+\right]=0\,,
\\\dm Q_{2+} -\dpl Q_{2-} +\left[\mc{A}_-,Q_{2+}\right]+\left[Q_{1-},P_+\right]+\left[P_-,Q_{1+}\right]+\left[Q_{2-},\mc{A}_+\right]=0\,.
\ea\ee
Here the first equation is automatically satisfied on the equations of motion \rf{fulleom}. 

The Pohlmeyer reduction procedure 
involves solving the equations of motion and the Virasoro constraints by introducing new variables  parametrizing the physical degrees of freedom. 
The equations of motion of the reduced theory are then the final three equations in the decomposed 
Maurer-Cartan equation \eqref{decompmc}.

Let us  briefly describe this   reduction
(for more details see section 6 of \ci{gt1}). 
The polar decomposition theorem implies firstly that we can always use a $G$ gauge transformation to set 
\be
P_+=p_{1+} T_1+p_{2+} T_2\,,
\ee 
and secondly write $P_-$ as follows
\be
P_-=p_{1-} g^{-1} T_1 g+p_{2-} g^{-1} T_2 g\,,
\ee
where $g$ is some element of $G= Sp\left(2,\,2\right)\x Sp\left(4\right)$ and $p_{1\pm}$ and $p_{2\pm}$ are functions of the worldsheet coordinates. $T_1$ and $T_2$ 
can be chosen  as follows
\be
\ba{c}
T_1=\fr{i}{2}\trm{ diag}\left(1,\, 1,\, -1,\, -1,\, 0,\,0,\, 0,\, 0\right)\,,
\\T_2=\fr{i}{2}\trm{ diag}\left(0,\, 0,\, 0,\, 0,\, 1,\, 1,\, -1,\, -1\right)\,.\ea\ee
These two elements span the maximal abelian subalgebra of $\mf{p}$. 
To solve the Virasoro constraints we may  then choose $p_+=p_{1+}=p_{2+}$ and similarly, $p_-=p_{1-}=p_{2-}$. Thus
\be\ba{c}\la{cur1}
P_+=p_{+} T\,,
\\P_-=p_{-} g^{-1} T g\,,
\ea\ee
where $T$ is defined as follows
\be
T=\fr{i}{2}\trm{ diag}\left(1,\, 1,\, -1,\, -1,\, 1,\, 1,\, -1,\, -1\right)\,.\ee
$T$ is an element of the maximal abelian subalgebra of $\mf{p}$. The group $H$ is then defined as
 the subgroup of $G$ which stabilizes $T$, that is $[h,\,T]=0$, $h\in H$. 

 One way of fixing the \kasym gauge is to project the fermionic currents onto the ``parallel space''
  \rf{kapsym} (see appendix \ref{PSU}), i.e.  
\be
Q_1=Q_1^\parallel, \hs{40pt} gQ_2 \gi=(gQ_2 \gi)^\parallel\,.
\ee
Substituting this into the fermionic equations of motion and noting that $[T,\mf{f}_{1,3}^\parallel]=2T \mf{f}_{1,3}^\parallel$, it is possible to see that solving the resulting equations implies
\be Q_{1-}=Q_{2+}=0\,. \ee 
The  equations of motion \eqref{fulleom} then become 
\be\ba{c}\la{boskafix}
\dpl P_- +\left[\mc{A}_+,P_-\right]=0\,,
\\\dm P_+ +\left[\mc{A}_-,P_+\right]=0\,.\ea\ee
Using the residual conformal diffeomorphism 
 symmetry it is always possible to set $p_\pm = \mu_\pm =$const., so that we get 
\be\ba{c}\la{bosfields1}
P_+=\mu_+ T\,,
\\P_-=\mu_- g^{-1}T g\,.
\ea\ee
 It should be noted that  if the sigma model 
 were  defined on  2-d Minkowski space then we could use a
  Lorentz transformation to set $\mu_+=\mu_-=\mu$ as was done in \ci{gt1} (and originally assumed 
   in \ci{pol}). However, 
  if we are interested in the case of the closed string 
  when the worldsheet is 
  $\mbb{R}\x S^1$ then  this 
  is not possible. 
  It will be  useful to define 
  the following  combination of $\mu_+$ and $\mu_-$, 
\be \mu=\sqrt{\mu_+\mu_-}\,.\ee
The  equations of motion \eqref{boskafix}  can be solved  as follows
\be\ba{c}\la{bosfields2}
\mc{A}_+=g^{-1}\dpl g+g^{-1}+g^{-1}A_+g\,,
\\\mc{A}_-=A_-\,.
\ea\ee
Here $A_+$ and $A_-$ are arbitrary fields taking values in the algebra $\mf{h}$
of $H$, i.e.
 $\left[A_\pm,T\right]=0$. Finally, we make the following redefinitions of the non-vanishing fermionic fields
\be\ba{c}\la{fermfields}
\YR=\fr{1}{\sqrt{\mu_+}}(Q_{1+})^{\parallel}\,,
\\\YL=\fr{1}{\sqrt{\mu_-}}(gQ_{2-}\gi)^{\parallel}\,.
\ea\ee

\subsection{Equations of motion and Lagrangian of  reduced theory\la{eqa}}

The equations of motion \eqref{fulleom} and the Virasoro constraints \eqref{vir}
have been solved by writing the original currents in terms of a new set of fields, $(g\,,A_\pm,\,\YR,\,\YL)$,  describing  only the physical degrees of freedom of the system. 
Substituting these into the second, third and fourth equations in \eqref{decompmc} we get the following set of equations of motion for the reduced theory
\be\ba{c}
\la{redeom}
\dm\left(\gi\dpl g+\gi A_+ g\right)-\dpl A_- +\left[A_-,\gi \dpl g + \gi A_+ g\right]\\\hs{140pt}=-\mu^2\left[\gi T g, T\right]- \mu\left[\gi\YL g, \YR\right]\,,
\\
\\D_-\YR=\mu\left[T,\gi \YL g\right]\,,\hspace{20pt}D_+\YL=\mu\left[T,g \YR \gi\right]\,,\hspace{20pt} D_\pm=\del_\pm+\left[A_\pm,\right]\,.
\ea\ee
These equations naturally have  $H\x H$  gauge symmetry,
\be\ba{c}\la{hxhgauge}
g\ra h^{-1}g\bar{h}\,,\hs{20pt}A_+\ra h^{-1}A_+h+h^{-1}\dpl h,\,\hs{20pt}A_-\ra \bar{h}^{-1}A_-\bar{h}+\bar{h}^{-1}\dm \bar{h}
\\\YR\ra\bar{h}^{-1}\YR \bar{h}\,,\hs{30pt}\YL\ra h^{-1}\YL h\,.\ea\ee
The factor of $H$ that corresponds to acting from the right on $g$
 arises as a subgroup from the original $G$ gauge freedom in the conformal-gauge  string theory. 
 The reason is that   once $P_+$ has been rotated to be proportional
  to $T$, it is still possible to perform further $G$ gauge transformations retaining this structure, as long as $g\in H$. The other factor of $H$, which corresponds to acting from the left on $g$
  arises because in defining the reduced theory field, $g$, there is an 
  ambiguity: 
   it is possible to let $g \ra h g$, where $h$ is an 
   arbitrary element of $H$, without changing that $P_-$ is proportional 
   to $g^{-1}T g$. Both of these gauge freedoms come about because $H$ is the stabilizer of $T$
    (i.e. $\left[h,\,T\right]=0$ for $h\in H$).

To be able to write down a sensible 
Lagrangian which leads to the equations of motion \eqref{redeom} we need to partially
 fix the $H\x H$ gauge symmetry to a  $H$ gauge symmetry. We can do this by demanding that
\be
\la{redgaugefields}\ba{c}
\tau\left(A_+\right)=\left(\gi\dpl g+\gi A_+ g-\fr{1}{2}\left[\left[T,\YR\right],\YR\right]\right)_\mf{h}\,,
\\\tau^{-1}\left(A_-\right)=\left(-\dm g \gi+g A_- \gi -\fr{1}{2}\left[\left[T,\YL\right],\YL\right]\right)_\mf{h}\,.
\ea\ee
Here $\tau$ (not to be confused with a time-like  world-sheet coordinate)
is a supertrace-preserving\foot{$\trm{STr}\left(\tau\left(u_1\right)
\tau\left(u_2\right)\right)=\trm{STr}\left(u_1u_2\right)$, \ $u_{1,2}\in\mf{h}$.} 
automorphism of the algebra $\mf{h}$. As discussed in \ci{gt1}, this partial 
gauge-fixing is always possible.\foot{Compared to  \ci{gt1}, we  choose to 
redefine $A_- \to \tau^{-1}\left(A_-\right)$.}
 The gauge symmetry is now reduced to the following 
asymmetric $H$ gauge symmetry,
\be\ba{c}\la{gaugetrans}
g\ra h^{-1}g\hat{\tau}\left(h\right)\,,\hs{20pt}A_+\ra h^{-1}A_+h+h^{-1}\dpl h,\,\hs{20pt}A_-\ra 
\hat{\tau}\left(h\right)^{-1}A_-\hat\tau\left(h\right)+\hat{\tau}\left(h\right)^{-1}\dm \hat\tau\left(h\right)
\\\YR\ra \hat{\tau}\left(h\right)^{-1}\YR \hat\tau\left(h\right)\,,\hs{30pt}\YL\ra h^{-1}\YL h\,,\ea\ee
where $\hat{\tau}$ is a lift of $\tau$ from $\mf{h}$ to $H$.

The equations of motion, \eqref{redeom}, and the gauge constraints, \eqref{redgaugefields},
 then follow from the following Lagrangian,\foot{
 The overall  coefficient  in  the reduced theory action should be  the same
  string tension that appears in
 the \adss string action.}
\begin{equation}\label{Ltot}
	\begin{split}
		\qquad \qquad L_{tot}&=L_{\rm gWZW}+\mu^2\,\Str(g^{-1}Tg T )\\
		&~~~+{\textstyle \frac{1}{2}}\mathrm{STr}\left(\Psi_{_L}\commut{T}{D_+\Psi_{_L}}+
		\Psi_{_R}\commut{T}{D_-\Psi_{_R}}\right) +\ \mu\,
		\mathrm{STr}\left(
		g^{-1}\Psi_{_L}g\Psi_{_R}\right)\,,
	\end{split}
\end{equation}
where $L_{\rm gWZW}$ is the Lagrangian  of  the asymmetrically  gauged  $G/H$ WZW model,
\begin{equation} \label{gauwzw}
	\begin{split}
		I_{\rm gWZW}  &= \int\frac{d^2\sigma}{4\pi}{\rm STr}(g^{-1}\partial_+g g^{-1}\partial_-g) - \int\frac{d^3\sigma}{12\pi}
		{\rm STr}(g^{-1}dg g^{-1}dg g^{-1}dg)\\
		&~~~+~\int \fr{d^2 \sigma}{2\pi} \Str \left(  A_+\,
		\partial_- g g^{-1} -
		A_- \,g^{-1}\partial_+ g-g^{-1} A_+ g  A_-  + \tau\left(A_+\right) A_- \right) \,.
	\end{split}
\end{equation}
This Lagrangian is invariant under the gauge transformations, 
\eqref{gaugetrans}, as expected. 


The reduced theory is thus the $G/H$ asymmetrically gauged WZW model with a
 gauge-invariant integrable potential and fermionic extension. For the case of the 
  superstring on $AdS_5 \x S^5$ we have $G=Sp\left(2,2\right)\x Sp\left(4\right)$ 
  and $H=[SU\left(2\right)]^4$. The embedding of these subgroups into $\Psu$ that 
  we use is discussed in appendix \ref{PSU}.

Let  us  stress that the equations of motion \rf{redeom}   obtained  directly 
from string theory equations after  solving the Virasoro  conditions in terms 
of new current variables do not ``know''  about the $\tau$-automorphism.  Thus 
the information contained  in  \rf{Ltot} with \rf{gauwzw}  that is relevant for string theory  
should also not depend  on $\tau$.  However, it is not clear a priori  (and seems seems unlikely) 
that the reduced theory actions with different choices of $\tau$ are 
 completely equivalent as 2-d quantum field theories. 

 In the sections 3 and 4  we shall 
 consider the case  of the symmetric  gauge fixing
when the automorphism $\tau$  is trivial, 
i.e. the reduced theory Lagrangian is given by \rf{Ltot},\rf{gauwzw} with $\tau =\bf 1 $.

\subsection{
 Vacua of the reduced theory\la{vac}}


The vacua  of  the reduced theory  may be  
 defined as  constant  solutions   which minimize 
 the potential $-\mu^2\,\Str(g^{-1}TgT)$ in  \rf{Ltot}. 
These are then 
\be\la{vuc}
g_{vac}=h_0\in H\,,\hs{30pt}h_0=\,\trm{const}\,.
\ee
Back in string theory all these vacua are equivalent  to the BMN vacuum.
As discussed above, when carrying out the reduction we initially have the 
equations \rf{redeom} with
$H\x H$ gauge symmetry, \eqref{hxhgauge}. 
We then use some of this gauge symmetry to fix the gauge fields as in \rf{redgaugefields}.

 Before this  partial gauge fixing it is always possible to choose the vacuum 
 in the equations \rf{redeom}   to be the identity, $g=\bf 1$:
 choices  of $g_{vac}=h_0\in H$ are gauge-equivalent. 
  After the  gauge fixing  needed to get a  Lagrangian set of equations of motion 
  this is no longer so: 
   we get a space of vacua \rf{vuc} that are not related by the 
   residual  
   $H$ gauge transformations. Still, they should 
    be effectively  equivalent 
   as far as the information relevant for string theory is concerned. 
   
  Let us emphasize that 
ultimately we are interested in observables of the 
 string theory.   We are only 
interested in  observables of  the
 reduced theory in
 the sense of what they say about the observables in 
 the  string theory.
 At the level of the equations of motion (i.e. classically) 
 it  is clear that the latter 
   should not depend on a particular 
  $H \x H \to H$ gauge-fixing. As the one-loop corrections are essentially determined  by 
 the equations of motion, this should  also be true at the 
  one-loop level (and should hopefully  be true in general). 

It is useful to note that expanding  the reduced theory action 
near  different vacua is related to using  different partial gauge-fixings
or different choices of $\tau$ in \rf{redgaugefields}.
Indeed, it is easy to see that  starting with  the action \rf{gauwzw} with $\tau= \bf 1$ 
and expanding it near $g= h_0$ is  equivalent to starting with \rf{gauwzw} with
the special choice of the automorphism  $\tau(u) = h_0^{-1} u h_0 $ 
and expanding it near $g= \bf 1$\la{page}.

As  was mentioned in  \ci{gt1},  there  is an 
  apparent  problem with expanding 
  the symmetrically gauged  ($\tau = \bf 1$)  action  \rf{Ltot} 
near the trivial vacuum, $g_{vac}=\mbf{1}$: 
the  $A_+A_- - g^{-1}A_+ g A_- $  part of the action, \eqref{gauwzw}, is then degenerate. 
This complication 
 may  be by-passed 
 by exploiting  the freedom to choose a different  gauging or
  a different vacuum in \rf{vuc} to expand around. 
 For example, one may  expand  the symmetrically gauged model  near 
\be \la{vacuum}g_{vac}=\left(\ba{cccc} 
i\si_i&\mbf{0}_2&\mbf{0}_2&\mbf{0}_2\\
\mbf{0}_2&i\si_j&\mbf{0}_2&\mbf{0}_2\\
\mbf{0}_2&\mbf{0}_2&i\si_k&\mbf{0}_2\\
\mbf{0}_2&\mbf{0}_2&\mbf{0}_2&i\si_l
\ea\right)\,,\ee
which is a constant matrix in 
 $H=\left[SU(2)\right]^4$. Here $\si_{1,2,3}$ are the Pauli matrices   and  
$i,\,j,\,k,\,l$ can take  any values 
$1,\,2,\,3$. It should be noted that these  choices are all related to each other by symmetric $H$ 
gauge transformations,
but are not equivalent to  $g_{vac}=\mbf{1}$.
Expanding  near this vacuum 
(combined   with an appropriate $H$ gauge fixing)
then   removes the  degeneracy.
This observation  may be useful for a  future study of the S-matrix
of  the reduced theory.



If we start with  the symmetrically gauged WZW model
we may parametrize $g$ in terms of eight bosonic scalar fields (after 
$H$ gauge fixing).\foot{Below  we will  not      
explicitly   relate $g$ 
to string coordinates  (we will always embed the string 
coordinates into $f$ and compute $g$ following the procedure outlined
 in section \ref{Class}).} 
We should  do this so that when these fields all vanish
we are left with $g_{vac}=h_0$, for constant $h_0\in H$\ (this includes $g_{vac}=\mbf{1}$ 
and also $g_{vac}$ as given in \rf{vacuum}).
At the level of the equations of motion \rf{redeom} 
these  choices are all related by $H\x H$ gauge transformations, but not 
by symmetric $H$ gauge 
transformations. Therefore, 
 there will be many solutions of  the symmetrically 
gauged WZW model, 
which are not related by $H$  gauge transformations,
but which correspond to  the same classical string solution (as they are 
related by a $H\x H$ gauge transformation, ignoring the gauge constraints). 
They may be distinguished by the vacuum they approach 
in the limit when the  string solution shrinks to a point.



In most  of this paper we will always look for classical solutions of the 
reduced theory such that they are solutions of the symmetrically gauged 
WZW model and have a vacuum limit that is related by a $H$ gauge
 transformation to \rf{vacuum}.\foot{In Appendix D we will consider the complex sine-Gordon
   and complex  sinh-Gordon models
 as truncated reduced theory models corresponding to the bosonic part of
  superstring theory on $AdS_3 \x S^3$. When  considering  these models we
   have already implicitly chosen a particular parametrization of $g$ in 
   terms of scalar fields, or,  equivalently, a particular embedding of the 
   string coordinates in $g$.
    This parametrization is 
   different from  the one used in the rest of the paper.
   }


\renewcommand{\theequation}{3.\arabic{equation}}
 \setcounter{equation}{0}

\section{Fluctuations near classical solution \\
from string theory  equations of motion \la{Fleom}}

In this section we shall  discuss fluctuations of the conformal-gauge string theory 
around  classical string solutions at the level of the equations of motion.
The underlying motivation is to 
compare one-loop  quantum corrections 
in string theory and  the reduced theory. 
Since the classical  
equations of the reduced theory are closely related  to the original conformal-gauge string equations 
(and their classical solutions are  in direct correspondence) 
the fluctuation spectra  near the respective solutions should also be closely 
related.

As discussed above, the string 
 theory equations can be written in terms of the current components built out of the 
  field $f\in \Psu$.
  Rather than fluctuating the currents directly 
  here  we will first  fluctuate  $f$ and 
  then consider how this affects the  equations of motion and the Maurer-Cartan equations
  for the currents.

    It is possible to parametrize $f$ in terms of fields that can be viewed as coordinates on $AdS_5 \x S^5$. The parametrization 
    that we use is discussed in appendix \ref{Coord}. Thus fluctuating $f$ is 
    equivalent to fluctuating these embedding coordinates. It is  still advantageous to  
    write the classical equations of motion in terms of the currents as then 
     the resulting fluctuation equations retain the algebra structure.

 One may   use the  Pohlmeyer reduction to simplify the fluctuations of the conformal-gauge string theory. 
Starting with a classical solution of the 
 reduced theory, if we are interested in the fluctuation spectrum 
   we do not need to  reconstruct the corresponding 
 classical form of $f$:
  we need  only to  know the corresponding   classical string theory
 currents. 
 We can  then substitute the reconstructed currents into the fluctuation equations of 
 the conformal-gauge string theory.

 This simplifies the fluctuation equations because
  in the Pohlmeyer reduction the $G$ gauge freedom of 
  the conformal-gauge string theory is used to rotate $P_+$ such that it is 
  proportional to $T$ (see section \ref{Class}). In terms of the embedding 
  coordinates on $AdS_5 \x S^5$, this is equivalent to choosing  the 
 coordinate system such that one of the directions of the worldsheet always 
 lies in a particular direction. The massless fluctuations, which are removed 
 via the Virasoro constraints, are the two fluctuations in the directions 
 along the worldsheet, while the physical fluctuations are those transverse to the worldsheet.
Since the Virasoro constraints are already solved in the reduced theory 
 it turns out to be much easier to isolate the physical fluctuations.

   Below we shall 
study  in detail a general class of  classical solutions
 living in an $AdS_2 \x S^2$ subspace of $AdS_5 \x S^5$ and consider
 the functional determinants of the operators acting on the physical
  fluctuations.\foot{Note that here while we are only considering classical 
  solutions living in $AdS_2 \x S^2$ we are fluctuating the canonical field  $f$ 
in all directions, including the fermionic directions.} 	For some special 
 solutions we will 
 see that the results
 will agree with the previously found  ones, such as for fluctuations near  the giant magnon  
 solution \cite{giantmagsc}.

The Lagrangian and the equations of motion for 
the conformal-gauge string 
 theory are given in \eqref{fulllag} and \eqref{fulleom} respectively.
We start with  a classical solution $f_0$  (with the  corresponding current  $J_0=\mc{A}_0+Q_{1\;0}+P_0+Q_{2\;0}$), and set 
\be
f=f_0 e^\xi\,,\hs{40pt}\xi\in\psu\,.
\ee
This should then be substituted into the classical equations of motion \rf{fulleom} and 
the Virasoro constraints \rf{vir}. The resulting equations are then 
expanded to first order in the fluctuation  field $\xi$. Since 
\be\ba{c} J=f^{-1} d f=f^{-1}_0 d f_0+\left[f^{-1}_0 d f_0,\xi\right]+d \xi+\mc{O}\left(\xi^2\right)+
\\\hs{30pt}=J_0+\left[J_0,\xi\right]+d\xi+\mc{O}\left(\xi^2\right)\ea\ee
is flat,  $dJ+J\wedge J=0$,  
 the fluctuation equations arising from the Maurer-Cartan 
equations will be satisfied automatically.

 We can split  the fluctuation field $\xi$ under the $\Z_4$ decomposition
\be
\xi=\xi_0+\xi_1+\xi_2+\xi_3\,.
\ee
Under the $\mbb{Z}_4$ grading $J$ decomposes as follows to first order in $\xi$, 
\be\la{decomp}\ba{c}
\mc{A}=\mc{A}_0+\left[\mc{A}_0,\xi_0\right]+\left[Q_{1\,0},\xi_3\right]+\left[P_0,\xi_2\right]+\left[Q_{2\,0},\xi_1\right]+d\xi_0
\\=\mc{A}_0+\de \mc{A}\,,
\\P=P_0+\left[\mc{A}_0,\xi_2\right]+\left[Q_{1\,0},\xi_1\right]+\left[P_0,\xi_0\right]+\left[Q_{2\,0},\xi_3\right]+d\xi_2
\\=P_0+\de P\,,
\\Q_1=Q_{1\,0}+\left[\mc{A}_0,\xi_1\right]+\left[Q_{1\,0},\xi_0\right]+\left[P_0,\xi_3\right]+\left[Q_{2\,0},\xi_2\right]+d\xi_1
\\=Q_{1\,0}+\de Q_1\,,
\\Q_2=Q_{2\,0}+\left[\mc{A}_0,\xi_3\right]+\left[Q_{1,\,0},\xi_2\right]+\left[P_0,\xi_1\right]+\left[Q_{2\,0},\xi_0\right]+d\xi_3
\\=Q_{2\,0}+\de Q_2\,.\ea\ee
Substituting these relations  back into \rf{fulleom} gives the equations of motion for
 the fluctuations $\xi$. These  need to be supplemented by the equations which
  arise from substituting \rf{decomp} back into the Virasoro conditions 
\rf{vir}  which  will give the constraint equations which remove the two massless
``longitudinal''  bosonic fluctuations.

\subsection{Fluctuations of  string equations
around  a classical solution of the \\ reduced theory\la{recon}}

The aim of this section is to determine  the fluctuations around a classical solution of the conformal-gauge string theory corresponding 
to a solution of  the reduced theory. 
Again, the  eventual goal is 
to show the equivalence between the  fluctuation spectrum and thus 
the one-loop corrections in string theory and in the reduced theory
in 
(cf. section \ref{Flact}).

 The strategy is to start with  a classical solution of the reduced theory and then
  reconstruct the classical currents of the conformal-gauge string theory. As already
   mentioned, we do not need to reconstruct the full classical solution of the 
   conformal-gauge string theory, $f_0$. 
 Given a classical solution of the reduced theory $g_0,\;A_{\pm 0},\;\YRo,\;\YLo$,
  the reconstructed currents of the  string theory solution are as follows
\be\ba{c}
P_{0+}=\mu_+ T\,,\hs{30pt}P_{0-}=\mu_- \gi_0 T g_0\,,
\\\mc{A}_{0+}=\gi_0\dpl g_0+\gi A_{+0} g_0\,,\hs{30pt}\mc{A}_{0-}=A_{-0}\,,
\\Q_{1\,0+}=\sr{\mu_+}\YRo\,,\hs{30pt}Q_{1\,0-}=0\,,
\\Q_{2\,0+}=0\,,\hs{30pt}Q_{2\,0-}=\sr{\mu_-}\gi_0\YLo g_0\,.
\ea \ee
Motivated by the comparison to the  reduced theory   let us 
 make the following redefinitions of the fermionic components of $\xi$
\be
\hat{\xi}_1=\fr{\xi_1}{\sr{\mu_-}}\,,\hs{30pt}\hat{\xi}_3=-\fr{g_0 \xi_3 \gi_0}{\sr{\mu_+}}\,.
\ee
We then fix the \kasym gauge by choosing $\hat{\xi}_1=\hat{\xi}_1^\parallel$ and $\hat{\xi}_3=\hat{\xi}_3^\parallel$.

 Substituting these formulae into \rf{fulleom} and \rf{vir}  gives the equations 
 of motion and constraint equations for the fluctuations. Here we will give 
 these equations  for the fluctuations with
vanishing classical fermionic content, i.e.
 $Q_{1\,0}=Q_{2\,0}=\YRo=\YLo=0$. 
  We will also assume that the classical solution of the reduced theory has 
  vanishing gauge fields, that is $A_{\pm 0}=0$. 
  It is possible to see that using the 
  $H$ gauge freedom and the fact that the current $A_0$ is flat,\foot{The flatness
   of $A_0$ comes from the equations of motion and thus implies
    that this is a statement that can only be made on-shell.} it 
    is always possible to choose the classical solution of the 
    reduced theory equations  \eqref{redeom} such that $A_{\pm 0}=0$. The fluctuation equations are then
\be\la{reconfluctfulleom}
\ba{c}
\dpm\xi_2+\dm\left[\gi_0\dpl g_0,\xi_2\right]+\mu^2\left[\left[\gi_0 T g_0,\xi_2\right],T\right]=0\,,
\\
\\\dm\hat{\xi}_1+\mu [T, [T,\gi_0 T,\hat{\xi}_3]g_0]]=0\,,
\\\dpl\hat{\xi}_3+\mu [T, [T,\gi_0 [T,\hat{\xi}_3g_0]]=0\,,
\ea
\ee
\be\la{reconfluctfullcon}
\ba{c}
\Str\left(T\left(\left[\gi_0\dpl g_0,\xi_2\right]+\dpl \xi_2\right)\right)=0\,,
\\\Str\left(\gi_0T g_0\dm \xi_2\right)=0\,.
\ea
\ee
Here we have used that  $T^2=-\fr{1}{4}\mbf{1}_8$ and the cyclicity of the supertrace to simplify the fluctuations of the  Virasoro constraints.

\subsection{Special case:  solutions  in  $AdS_2 \x S^2$ subspace  of  $AdS_5 \x S^5$\la{2x2}}

Let us  consider  a particular case of the 
classical string solutions in a $AdS_2 \x S^2$ subspace of $AdS_5 \x S^5$ and the 
 corresponding classical solutions in the reduced theory
(see also  appendix \ref{coord2x2} for details). 
We will see that the  resulting functional determinants, which determine 
the one-loop corrections 
match the corresponding functional determinants in 
the reduced theory computed in section \ref{act2x2}.

 There are many interesting string solutions which live in $AdS_2 \x S^2$,
  and using the results of this section it may be possible to better
   understand the one-loop corrections to their energies. 
   The simplest are the ones that 
effectively live in $\mbb{R}_t\x S^1$, that is the point-like orbiting
  string (i.e. the geodesic corresponding to the BMN vacuum state) and the (unstable)
 static string wrapped on a big circle of $S^5$. 
 There are no homogeneous string solutions  in $AdS_2 \x S^2$ apart from these two 
  special  cases, but  there are many other  simple configurations:
   pulsating strings, folded strings and finite-size  magnons
   (see \ci{puls,finsizemag1,finsizemag2,okamura,hosv} and 
   references therein). One of the limits of the finite-size magnon is the giant magnon \ci{hm},
    for which the one-loop correction was  shown to vanish 
   \cite{giantmagsc}. We shall compare  our results against the expressions 
    in this paper in appendix \ref{Check}
     and show  that they agree.

 As discussed in appendix \ref{coord2x2}, for the  bosonic solutions  in 
 $AdS_2 \x S^2$ we can consider the following element of $G$ as the field used to 
 parametrize $P_-$ in the reduction procedure \rf{cur1}
\be\la{gA2S2}
g_0=\left(\ba{cc}
g_A&\mbf{0}_4
\\\mbf{0}_4&g_S\ea\right)\,,
\ee
\bd
g_A=\left(\ba{cccc}
i\cosh\phi_A&0&0&\sinh\phi_A
\\0&-i\cosh\phi_A&\sinh\phi_A&0
\\0&\sinh\phi_A&i\cosh\phi_A&0
\\\sinh\phi_A&0&0&-i\cosh\phi_A
\ea\right)\,,\ed
\bd
g_S=\left(\ba{cccc}
i\cos\phi_S&0&0&i\sin\phi_S
\\0&-i\cos\phi_S&i\sin\phi_S&0
\\0&i\sin\phi_S&i\cos\phi_S&0
\\i\sin\phi_S&0&0&-i\cos\phi_S
\ea\right)\,,\ed
where $\phi_A$ and $\phi_S$ satisfy 
\be\ba{c} \la{sinn}
\dpm\phi_A+\fr{\mu^2}{2}\sinh2\phi_A=0\,,
\\\dpm\phi_S+\fr{\mu^2}{2}\sin2\phi_S=0 \,.
\ea\ee

For this configuration $A_{\pm 0}=\YRo=\YLo=0$ (see appendix \ref{coord2x2}). It should be 
noted that as $\gi_0 \dpl g_0 \in \mf{m}$ and $\dm g_0 \gi_0\in \mf{m}$ (as defined in appendix \ref{PSU}),
 $A_{\pm 0}=0$ is a consistent solution for the gauge fields. This  configuration
  satisfies 
 the classical equations of motion 
 of the reduced theory, provided 
  $\phi_A$ and $\phi_S$ satisfy the above sinh-Gordon and sine-Gordon equations.

\subsubsection{Fluctuations near $AdS_2 \times S^2$ solution in 
 string theory}

The fluctuations around the corresponding  solution in the conformal-gauge string theory
can be found   following the method in section \ref{recon}. That is, we start from
 the reduced theory classical solution, reconstruct the classical currents of the 
 string theory, and substitute these into the equations of motion for the fluctuations
  of the field  $f$.

 Substituting \rf{gA2S2} into \rf{reconfluctfulleom} and considering the components of the resulting matrix equations the following fluctuation equations arise.
 In the bosonic $AdS_5$ sector we get
\be\la{fullA2fluct1}
\dpm z_i +\mu^2 \cosh 2\phi_A\;z_i\equiv  \mc{O}_1 \;z_i=0\,,\hs{30pt}i=1,\,2,\,3
\ee
and one copy of the following set of coupled equations
\be\la{fullA2fluct2coup}
\dpm z_4+\mu^2\cosh2\phi_{A}\;z_4-2\,\dpl \phi_A \dm z_5=0\,,
\ee
\be\la{fullA2fluctmassless}
\dm\left(\dpl z_5-2\,\dpl\phi_A\,z_4\right)=0\,.
\ee
Here $z_1,\;z_2,\;z_3,\;z_4,\;z_5$ are the five components of $\xi_2$ in the $AdS_5$ sector.

We still need to impose the constraints on the fluctuations arising from fluctuating the Virasoro constraints, \rf{reconfluctfullcon}. For the present
  configuration the $AdS_5$ and $S^5$ sectors are decoupled and thus in the Virasoro constraints we can split up the supertrace into traces over the 
  two sectors  and demand that they both vanish separately. The following constraints then arise for the fluctuations in the $AdS_5$ sector,
\be\la{vir2x2}
\ba{c}
\dpl z_5-2\,\dpl\phi_A\,z_4=0\,,
\\\dm z_5-\tanh2\phi_A\,\dm z_4=0\,.
\ea
\ee
It is possible to see that both \eqref{fullA2fluct2coup} and \eqref{fullA2fluctmassless} are implied by \eqref{vir2x2}. 
This coupled first order system is equivalent to the second order system, obtained by eliminating $z_4$ or $z_5$. Thus the relevant fluctuation 
operator can be found by either eliminating $z_4$ or $z_5$ from \eqref{vir2x2} or by just 
considering the coupled first order operator. These should lead to  the same functional determinant.

Here we choose to eliminate $z_5$, resulting in the following equation for the fourth 
physical bosonic fluctuation  $z_4$ in the $AdS_5$ sector (the other three are given by \rf{fullA2fluct1})
\be\la{fullA2fluct2}
\dpm z_4+\mu^2\cosh2\phi_A\;z_4-2 \,\tanh2\phi_A\,\dpl \phi_A\dm z_4\equiv \mc{O}_2\;z_4=0\,.
\ee 
Let us show that  the determinants of the two operators,
\be\ba{c}
\mc{O}_1=\dpm +\mu^2 \cosh 2\phi_A
\ea\ee
and
\be\ba{c}
\mc{O}_2=\dpm+\mu^2\cosh2\phi_A-2 \,\tanh2\phi_A\,\dpl \phi_A\dm
\ea\ee
are equal. Defining
\be
V_A\equiv \mu^2 \cosh 2\phi_A\,,
\ee
we have 
\be\la{A2fluct2llt}\ba{c}
\mc{O}_1=\dpm +V_A\hs{30pt} \trm{and}\hs{30pt}\mc{O}_2=\left(V_A\dpl\right)\left(V_A^{-1}\dm\right)+V_A\,.
\ea\ee
Considering the product 
\be
\left(\ba{cc}V_A &0\\0&V_A^{-1} \ea\right)\left(\ba{cc}\dpl&-1\\V_A&\dm\ea\right)=
\left(\ba{cc}V_A\dpl&-V_A\\1&V_A^{-1}\dm\ea\right)\,
\ee
and taking the determinant\foot{For a matrix of operators we have\bd\trm{det}
\left(\ba{cc}A&B\\C&D\ea\right)=\trm{det}\left(AD-ACA^{-1}B\right)=\trm{det}\left(DA-CA^{-1}BA\right)\,.\ed} on 
both sides we immediately see that 
\be \trm{det}\,\mc{O}_1=\trm{det}\,\mc{O}_2   \ . \ee

 Therefore, the contribution of the $AdS_5$  sector to the one-loop correction is given by 
the four copies of the  determinant of $\mc{O}_1$, i.e. 
\be\la{A2fluctfull}
4 \ln \det ( \dpm +\mu^2 \cosh 2\phi_A)  \ . \ee
In the bosonic $S^5$ sector the story is  the same,
  with  $V_A \to V_S=\mu^2 \cos2 \phi_S$. 
 The contribution   of this sector 
 is then  given by 
\be\la{S2fluctfull}
4 \ln \det (\dpm +\mu^2 \cos 2\phi_S)\,.
\ee
For the fermionic fluctuations we get the following sets of coupled equations
\be\la{A2S2fermfluct}
\ba{cc}
\dm \vt_{i} +\mu\,\cos\phi_S\cosh\phi_A\,{\tvt}_{i}+\mu\,\sin\phi_S\sinh\phi_A\,{\tvt}_{i+1}=0\,,&
\\\dpl {\tvt}_{i}-\mu\,\cos\phi_S\cosh\phi_A\,\vt_{i}+\mu\,\sin\phi_S\sinh\phi_A\,\vt_{i+1}=0\,,&\hs{30pt}i=1,\,3,\,5,\,7
\\\dm \vt_{i+1}+\mu\,\cos\phi_S\cosh\phi_A\,{\tvt}_{i+1}-\mu\,\sin\phi_S\sinh\phi_A\,{\tvt}_{i}=0\,,&\hs{30pt}
\\\dpl {\tvt}_{i+1}-\mu\,\cos\phi_S\cosh\phi_A\,\vt_{i+1}-\mu\,\sin\phi_S\sinh\phi_A\,\vt_{i}=0\,.&
\ea
\ee
Here the anticommuting functions $\vt_k$  are components of $\hat{\xi}_1\in \mf{f}^\parallel_1$ and ${\tvt}_k$ 
 are components of $\hat{\xi}_3\in\mf{f}^\parallel_3$. The 16 coupled first 
 order equations can be rearranged into 8 coupled second order equations describing the
 expected
  8 fermionic degrees of freedom.

 In appendix \ref{Check} the results of this section are 
applied to the case of 
 the giant magnon classical solution \ci{hm}
 and shown to agree   with  \cite{giantmagsc}, where the
  one-loop correction to the energy was computed by fluctuating the embedding coordinates.

\subsubsection{Solutions in $\mbb{R}_t\x S^1$}

There are two special  string solutions that live in $\mbb{R}_t\x S^1$ with  $\mbb{R}_t $ from  $AdS_5
$ and $S^1$ from  $ S^5$. 
These are  (i) the (supersymmetric) point-like orbiting string,
 $t=\ka\ta,\;\thet=\ka\ta$, and  (ii)  the (unstable) static wound  
 closed  string, $t=k\ta,\;\thet=k\si,\;k\in\mbb{Z}$
 ($k$ is the winding number). Here $t$ and $
 \thet$ are the coordinates in $AdS_5$ and $ S^5$ as defined in Appedix \ref{Coord}. 

 The reduced theory solutions corresponding to these 
two string solutions are the constant solutions of the sinh-Gordon and sine-Gordon equations. For the
 sinh-Gordon 
one the only constant solution is $\phi_A=0$. For the sine-Gordon equation  the constant solutions are 
$\phi_S=\fr{n\pi}{2}$, $n\in \mbb{Z}$. These break down into
 two distinct types, either $\phi_S=n\pi$ or $\phi_S=n \pi+\fr{\pi}{2}$, which correspond to minima 
 and maxima of the potential, $\mu^2 \cos 2\phi_S$; these lead to 
   stable and unstable solutions respectively.

The reduced theory solution $\phi_A=\phi_S=0$  gives
the point-like  string in $\mbb{R}_t\x S^1$
 in  string theory, with $\mu=\kappa$.  Thus  
 a stable vacuum solution of the reduced theory corresponds 
  to the stable BMN vacuum solution of the conformal-gauge string theory. 
 The bosonic and fermionic fluctuation equations are then the  familiar one 
\be
\dpm \ze_i+\mu^2\ze_i=0\,,\hs{40pt}i=1,\,\ldots,\,8\,.
\ee
\be
\dpm \vt_i+\mu^2\vt_i=0\,,\hs{40pt}i=1,\,\ldots,\,8\,.
\ee
For the  static  string  wrapped on $S^1 $ in $S^5$
the corresponding 
 reduced theory solution is  
  $\phi_A=0,\;\phi_S=\fr{\pi}{2}$,  
 with $\mu=k$. As expected,   an  unstable solution in the 
 reduced theory gives rise to an unstable solution in  string theory. The bosonic $AdS_5$  and $S^5$ fluctuation equations are
 respectively 
\be
\dpm \ze_i+\mu^2\ze_i=0\,,\hs{40pt}i=1,\,\ldots,\,4\,.
\ee
\be
\dpm \ze_i-\mu^2\ze_i=0\,,\hs{40pt}i=5,\,\ldots,\,8\,.
\ee
The fermionic fluctuation equations are
\be
\dpm \vt_i=0\,,\hs{40pt}i=1,\,\ldots,\,8\,.
\ee
For both the above solutions the fluctuation spectra   computed in the reduced theory and 
directly in the string theory (using, e.g., the embedding coordinates) 
match, and thus the one-loop  partition functions  also  match, providing a simple 
check of our general claim.


\renewcommand{\theequation}{4.\arabic{equation}}
 \setcounter{equation}{0}

\section{Fluctuations near classical solution \\
from  the action of the reduced theory\la{Flact}}

In this section we will  investigate the quadratic 
fluctuations in  the reduced theory action expanded around classical solutions.
Again, the  aim  is to see whether the sum of logarithms of the 
functional determinants which gives
 the one-loop  partition  function of 
 the reduced theory is the same as  in the conformal-gauge string theory
 expanded near the corresponding solution.

While we will  not prove in general that the one-loop partition functions match, 
we shall demonstrate the  equivalence for certain classes of classical solutions. These include  solutions which live in an $AdS_2 \x S^2$ subspace of $AdS_5 \x S^5$ and also homogeneous solutions of the conformal-gauge string theory.

We shall  parametrize the  basic variable of the reduced theory
  $g\in G $ as follows
\be
g=g_0e^\h\,, \hs{30pt}\h\in\mf{g}\,,
\ee
where  $\eta$ is the  fluctuation field.

Under the $\mbb{Z}_2$ decomposition discussed in appendix \ref{PSU} we have 
  $\h=\h^\parallel+\h^\perp$ where $\h^\parallel\in \mf{m}$ and $\h^\perp \in\mf{h}$. As 
  the physical bosonic fluctuations should be those 
  corresponding to the coset $G/H$
part,  we will take them to be the components of $\h^\parallel$.\foot{One may think of
 these fluctuations as corresponding to cartesian coordinates
(as opposed to radii 
and angles), cf. \rf{cart}.}
  As expected,  there are eight independent  
  components of the  bosonic fluctuation field  $\h^\parallel$.

The fields  $\h^\perp$ and  the fluctuations of the gauge fields, $\de A_\pm\in \mf{h}$,  will, in general, 
be coupled to $\h^\parallel$. 
To isolate the physical fluctuations the $H$ gauge needs to be fixed. We will
 always choose to fix the gauge on $\h^\perp$ and $\de A_\pm$, understanding that 
 the components of $\h^\parallel$ should  be  the physical fluctuations.

An  evidence that the components of $\h^\parallel$ are the 
physical fluctuations is that in the quadratic fluctuation Lagrangian, 
\rf{quafluc}, the kinetic term 
is given by \bd\Str(\dpl \h \dm \h)=\Str(\dpl \h^\parallel \dm 
\h^\parallel)+\Str(\dpl \h^\perp \dm \h^\perp)\,.\ed 
Expressing 
 $\h$ in terms of the 
component fields gives kinetic terms
 with the correct sign for the fields in $\h^\parallel$, but the 
 wrong sign for the some of the fields in $\h^\perp$.

It should be noted that under the $H$ gauge transformations we have
\bd\ba{c}
\eta\ra h^{-1}\eta h\ \  \Ra\ \ 
 \eta^{\parallel}\ra h^{-1}\eta^{\parallel} h, \hs{20pt} \eta^{\perp}\ra h^{-1}\eta^{\perp} h\,.
\ea\ed
Therefore, the components of $\h^\parallel$ and $\h^\perp$ cannot mix under these transformations. 






\subsection{Expansion of the reduced theory action\la{actred}}

The reduced theory action found in \ci{gt1,mikhshaf}
is a particular 
fermionic extension  of the   $G/H$ left-right symmetrically gauged WZW model
 with a $H$ gauge invariant integrable potential (for its detailed discussion see also 
\ci{rtfin}). 
In the  case of the $AdS_5 \x S^5$ superstring 
  we have $G=Sp\left(2,2\right)\x Sp\left(4\right)$ and $H=[SU\left(2\right)]^4$. 
  The embedding of these subgroups into $\Psu$ that we use is discussed in appendix
   \ref{PSU}. The Lagrangian and the equations of motion for this theory were given 
   in \eqref{Ltot} and \eqref{redeom} respectively.

 We consider the fluctuations around a classical solution, $g_0$, $A_{\pm 0}$, $\Psi_{_R0}$, $\Psi_{_L0}$, as follows 
\be \ba{c}
	g=g_0 e^\h = g_0(1+\h +\ha \h^2 +\mc{O}(\h^3))  \,,
	\\A_+=A_{+0}+\de A_+ \,,~~~~~~ A_-=A_{-0}+\de A_- \,,
	\\\Psi_{_R}=\Psi_{_R0}+\de \Psi_{_R} \,, ~~~~~~ \Psi_{_L}=\Psi_{_L0}+\de \Psi_{_L}\,.
\ea\ee
Below  we will only consider classical solutions with vanishing 
fermionic content, i.e. 
 $\Psi_{_R0}$ and $\Psi_{_L0}$ will be set to zero. 
 The quadratic fluctuation part of the Lagrangian \eqref{Ltot} is then
\begin{equation} \label{quafluc}
	\begin{split}
	 L_{quad}&={\rm STr} \Bigg[ \ha \dpl \h \dm \h + \ha \left( \h \dm \h -\dm \h \h \right) g_0^{-1}\! \dpl g_0 +\de A_+ g_0 \dm \h \gi _0  
	- \ha A_{+0}g_0 \dm \h \h \gi _0  \\ 
	 &~~+ \ha A_{+0} g_0 \h \dm \h \gi _0 + \de A_- \h \gi _0 \! \dpl g_0  -\de A_-\gi_0  \! \dpl g_0 \h -\de A_- \dpl \h   + A_{-0} \h \gi _0 \dpl g_0 \h \\  
	&~~+ \ha A_{-0}\h \dpl \h -\ha A_{-0}\h ^2 \gi _0 \! \dpl g_0   -\ha A_{-0}\gi_0  \dpl g_0 \h ^2 -\ha A_{-0}\dpl \h \h  
	+\de A_+ \de A_- \\
   &~~-\ha \h ^2 \gi _0 A_{+0} gA_{-0} -\ha \gi _0 A_{+0} g_0 \h ^2 A_{-0} +
   \h \gi _0 \de A_+ g_0 A_{-0}   + \h \gi _0 A_{+0}g_0 \h A_{-0} \\ &~~
    +\h \gi _0 A_{+0}g_0 \de A_-     - \gi _0  \de A_+ g_0 \h A_{-0} -\gi _0\de A_+
    g_0\de A_-  -\gi _0A_{+0} g_0 \h \de A_- \\ &~~ +\mu ^2\big( \ha \h ^2 \gi _0 T
     g_0 T + \ha \gi _0 T g_0 \h ^2 T-\h \gi _0 T g_0 \h T \big)    
	 \\ &~~+ \ha \de \Psi_{_R} \commut{T}{\dm \de \Psi_{_R}+
	\commut{A_{-0}}{\de \Psi_{_R}}} 
	+\ha \de \Psi_{_L} \commut{T}{\dpl \de \Psi_{_L}+\commut{A_{+0}}{\de \Psi_{_L}}}
	 +\mu \gi_0 \de \Psi_{_L} g_0\de \Psi_{_R} \Bigg] \,. 
	\end{split}
\end{equation}

For $\Psi_{_{R,L} 0}=0$ the  bosonic  and fermionic fluctuations decouple at quadratic order;
the  fermionic sector  describes only 
the physical fermionic degrees of freedom, that is the sixteen real 
anticommuting fields parametrizing $\de \YR$ and $\de \YL$. Therefore, 
to determine the operator which acts on the fermions, we can simply extract it from 
 the equations of motion for the fermionic fluctuations.

To isolate the physical bosonic fluctuations an  $H$ gauge needs to be fixed
(or 
the unphysical fluctuations integrated out). 
The  $H$ gauge symmetry acts  as follows (see \eqref{gaugetrans})
\be\ba{c}
g_0 e^\h= g\ra h^{-1} g h=h^{-1}g_0h \,e^{h^{-1}\h h}\,,
\\A_{\pm 0}+\de A_\pm=A_\pm\ra h^{-1}A_\pm h+h^{-1}\partial_\pm h=h^{-1}A_{\pm 0} h+h^{-1}\partial_\pm h+h^{-1}\de A_\pm h\,,
\\ \YRo+\de\YR=\YR\ra h^{-1}\YR h= h^{-1}\YRo h+ h^{-1}\de\YR h\,,
\\\YLo+\de \YL =\YL\ra h^{-1}\YL h= h^{-1}\YLo h+ h^{-1}\de\YL h\,.
\ea\ee 
These relations determine the transformations of the 
fluctuation fields and allow to fix an $H$ gauge.

\subsection{Fluctuations near solutions in $AdS_2 \x S^2$ subspace of  $AdS_5 \x S^5$\la{act2x2}}

Let us consider the particular case 
of the expansion near the reduced theory 
 solutions corresponding to the string theory 
 solutions in  $AdS_2 \x S^2$ subspace of $AdS_5 \x S^5$. As 
 discussed in appendix \ref{coord2x2}, such reduced theory solutions
 can be parametrized as in \rf{gA2S2},\rf{sinn}.
To fix the  $H$ gauge let us first note that we can always
 write $\h^\perp$ and $\de A_{\pm}$ as
\bd
\h^\perp=\left(\ba{cccc}
u_1+\si_3u_2^*\si_3&\mbf{0}_2&\mbf{0}_2&\mbf{0}_2
\\\mbf{0}_2&u_1^*-\si_3u_2\si_3&\mbf{0}_2&\mbf{0}_2
\\\mbf{0}_2&\mbf{0}_2&u_3-\si_3u_4\si_3&\mbf{0}_2
\\\mbf{0}_2&\mbf{0}_2&\mbf{0}_2&u^*_3+\si_3u^*_4\si_3
\ea\right)\,,\hs{20pt} \si_3=\left(\ba{cc}1&0\\0&-1\ea\right)\,,\ed
\bd \la{tak}
\de A_{+}=\left(\ba{cccc}
\rm{a}_{1+}-\rm{a}_{2+}^*&\mbf{0}_2&\mbf{0}_2&\mbf{0}_2
\\\mbf{0}_2&\rm{a}^*_{1+}+\rm{a}_{2+}&\mbf{0}_2&\mbf{0}_2
\\\mbf{0}_2&\mbf{0}_2&\rm{a}_{3+}+\rm{a}_{4+}&\mbf{0}_2
\\\mbf{0}_2&\mbf{0}_2&\mbf{0}_2&\rm{a}^*_{3+}-\rm{a}^*_{4+}
\ea\right)\,,\\
\ed \bd
\de A_-=\left(\ba{cccc}
\rm{a}_{1-}+\si_3\rm{a}_{2-}^*\si_3&\mbf{0}_2&\mbf{0}_2&\mbf{0}_2
\\\mbf{0}_2&\rm{a}_{1-}^*-\si_3\rm{a}_{2-}\si_3&\mbf{0}_2&\mbf{0}_2
\\\mbf{0}_2&\mbf{0}_2&\rm{a}_{3-}-\si_3\rm{a}_{4-}\si_3&\mbf{0}_2
\\\mbf{0}_2&\mbf{0}_2&\mbf{0}_2&\rm{a}^*_{3-}+\si_3\rm{a}^*_{4-}\si_3
\ea\right)\,.\ed
Here $u_i$ and $\rm{a}_{i\pm}$\,, \  ($i=1,\,2,\,3,\,4$),  are all elements of $\mf{su}(2)$. The $H$ gauge is then partially fixed by choosing
\be\la{gaugefix2x2a}
(\dm u_2-\rm{a}_{2-})\dpl\phi_A+\fr{1}{2}\dm(\rm{a}_{2+}\sinh 2\phi_A)=0\,,
\ee
\be\la{gaugefix2x2b}
(\dm u_4-\rm{a}_{4-})\dpl\phi_S+\fr{1}{2}\dm(\rm{a}_{4+}\sin 2\phi_S)=0\,.
\ee
This $H$ gauge choice fixes 6 of the 12 degrees of freedom of the $H$ gauge symmetry.
The reason for choosing this gauge is that when the classical solution and the gauge-fixing conditions
 are substituted into the quadratic fluctuation Lagrangian, \eqref{quafluc}, the physical 
 fluctuation fields, $\eta^\parallel$, decouple from the remaining 
  unphysical fluctuation fields.

 
Now that the physical fluctuations have been decoupled we should be able to use the remaining $H$ gauge symmetry
to ensure that the sector of the Lagrangian containing the unphysical bosonic fluctuations does not produce a non-trivial contribution.
 %

 We are then  left with the decoupled physical bosonic fluctuations. 
We may introduce the components of $\eta^\parallel$ as follows
\be \la{cart}\eta^\parallel=\left(\ba{cc}\eta_A^\parallel&0\\0&\eta_S^\parallel\ea\right)\,,\ee
\bd\eta_A= \fr{1}{\sqrt{2}}\left(
	\begin{array}{cccc}
		0 & 0 & \zeta_{1}+i\zeta_{2}& \zeta_{3}+i\zeta_{4}\\
		0 & 0 & \zeta_{3}-i\zeta_{4}& -\zeta_{1}+i\zeta_{2}\\
		\zeta_{1}-i\zeta_{2}& \zeta_{3}+i\zeta_{4}& 0 & 0 \\
		\zeta_{3}-i\zeta_{4}& -\zeta_{1}-i\zeta_{2}& 0 & 0  
	\end{array}
	\right)\,,\ed\bd
	\eta_S= \fr{1}{\sqrt{2}}\left(
	\begin{array}{cccc}
		0 & 0 & \zeta_{5}+i\zeta_{6}& \zeta_{7}+i\zeta_{8}\\
		0 & 0 & \zeta_{7}-i\zeta_{8}& -\zeta_{5}+i\zeta_{6}\\
		\zeta_{5}-i\zeta_{6}& \zeta_{7}+i\zeta_{8}& 0 & 0 \\
		\zeta_{7}-i\zeta_{8}& -\zeta_{5}-i\zeta_{6}& 0 & 0  
	\end{array}
	\right)\,.
\ed
The corresponding part of the fluctuation  Lagrangian  is then
\be
L_{b}=-\sum_{i=1}^4\zeta_i(\dpm+\mu^2\cosh2\phi_A)\zeta_i- \sum_{i=5}^8\zeta_i(\dpm+\mu^2\cos2\phi_S)\zeta_i\,.
\ee
Then  the resulting bosonic part of the one-loop partition function is 
given by 
\be   \Big(\Big[\det ( \dpm+\mu^2\cosh2\phi_A)  \ \det (\dpm+\mu^2\cos2\phi_S)\Big]^4 \Big)^{-1/2} \ . 
\la{zee}
\ee
This is the same result  as was  found for  the fluctuations of 
the conformal-gauge string theory around the classical solutions in
 $AdS_2 \x S^2$ subspace of $AdS_5\x S^5$ 
in section \ref{2x2} (see \eqref{A2fluctfull} and \eqref{S2fluctfull}).

 As was already  mentioned, to determine the fermionic fluctuation operator 
  we may  just use 
  the equations of motion arising from varying 
  the quadratic fluctuation Lagrangian, \eqref{quafluc}. It is easy to see that 
  these give the same  operator as  found for the conformal-gauge
  string theory fermionic fluctuations, see \eqref{A2S2fermfluct}.

  We  conclude that for the classical solutions of $AdS_5 \x S^5$ string
   theory  localized  in $AdS_2 \x S^2$, the one-loop 
   partition functions computed in the  string theory and in the  reduced theory are the same. 


\subsection{Homogeneous string solutions\la{acthom}}


Let us  now consider another  class of  $AdS_5 \x S^5$ string solutions --    
``homogeneous solutions'' \ci{ft3,ft2,fpt,ptt,btz} --
for which the string has rigid shape  and 
  for which one  can arrange to have 
 the coefficients in the quadratic fluctuation Lagrangian to be 
constant. In this case the  determinants of 
 the operators which enter  the one-loop partition function 
 are  expressed in terms of the characteristic frequencies which 
 are relatively simple to calculate and compare between the conformal-gauge 
 string theory and the reduced theory.

 Our  approach will be to start with  a homogeneous solution of the
 conformal-gauge string theory and construct the corresponding 
  field  $f$  using the parametrization of $PSU(2,2|4)$ 
 in terms of the embedding coordinates as described in appendix \ref{Coord}. 
 Then  the classical solution of the reduced theory will be found 
  following the reduction procedure 
  outlined in section \ref{Class}. 
   Since  in the process of the reduction the natural $H\x H$ gauge symmetry 
  of the string equations of motion 
  is partially fixed to a 
  $H$ gauge symmetry, the solution of the reduced theory
  will correspond to the string theory solution in  this partial gauge.

We use the $H$ gauge symmetry to choose the classical 
solution of the reduced theory such that $\gi_0 \del_\pm 
g_0$ and $\gi_0 T g_0$ are constant. This is possible for the 
homogeneous solutions that we consider below 
(and should be possible in general).
The reason for choosing this gauge
 is to help construct a Lagrangian for the  physical fluctuations
 which has  constant coefficients.

The quadratic fluctuation Lagrangian, \eqref{quafluc}, can  then be   used to 
find  the characteristic frequencies of fluctuations around the reduced theory 
 solution. 
 We will see that is possible to   choose  the $H$ gauge on the fluctuations  so that 
the coefficients in the quadratic fluctuation Lagrangian  for the eight 
bosonic and eight fermionic physical fluctuations are all  
constant. It is then easy to
 compute 
the corresponding fluctuation frequencies.
The  resulting  fluctuation frequencies around the classical solutions of the reduced theory 
 will be  shown
 to match the previously  found frequencies of fluctuations around the  homogeneous 
 solutions in  string theory.

 
\subsubsection{Homogeneous string solution in $\mbb{R}_t \times S^3$}

One example of a simple   string theory solution we shall consider here 
 is the rigid circular 
two-spin string  on  $ S^3$ in  $ S^5 $  discussed in \ci{ft3,art,ptt,btz}.
Using the embedding coordinates  in appendix \ref{Coord},
i.e.  $Y_{M}$ $(M=-1,\;0,\ldots,\;4)$ of $\mbb{R}^{4,2}$ for the $AdS_5$ part 
and $X_{I}$ $(I=1,\;2,\;\ldots,\;6)$ of $\mbb{R}^6$ for the $S^5$ part, 
this bosonic  string  solution is
\be\la{rxs3hom}
\ba{c}
Y_0+iY_{-1}=e^{i\ka \ta}\,, ~~~~~~ Y_1=Y_2=Y_3=Y_4=0\,,
\\X_1+iX_{2}=\fr{1}{\sqrt{2}}e^{i\om \ta +im \si}\,, ~~~~~~ X_3+iX_{4}=\fr{1}{\sqrt{2}}e^{i\om \ta -im \si}\,.~~~~~~X_5=X_6=0\,, 
\\
\ea\ee
The Virasoro constraints imply that the three parameters, $\ka$, $\om$ and $m$, are related by 
\be \la{rell}   \ka ^2=m^2+\om^2  \ . \ee
Using the parametrizations discussed in appendix B
we obtain the corresponding bosonic coset element $f$, 
\begin{gather}
\ba{c}
	f=\left(
	\begin{array}{cc}
		f_A& \mbf{0}_4  \\
		\mbf{0}_4& f_S
	\end{array}
	\right) \,,
\\	f_A=\left(
	\begin{array}{cccc}
		e^{\fr{i\ka \ta}{2}} & 0 &0 & 0 \\
		0 & e^{\fr{i\ka \ta}{2}} &0 & 0 \\
		0 & 0 &e^{-\fr{i\ka \ta}{2}}&0\\
		0 & 0 &0 &e^{-\fr{i\ka \ta}{2}} 
	\end{array}
	\right) \,,
\\	f_S=\left(
	\begin{array}{cccc}
		\fr{1}{\sqrt{2}} & 0 &\fr{i}{2}e^{-i\om \ta +i m\si} &-\fr{i}{2}e^{-i\om \ta -i m\si}  \\
		0 &\fr{1}{\sqrt{2}} &\fr{i}{2}e^{i\om \ta +i m\si}  &\fr{i}{2}e^{i\om \ta -im\si}  \\
		\fr{i}{2}e^{i\om \ta -im\si} & \fr{i}{2}e^{-i\om \ta -im\si}  &\fr{1}{\sqrt{2}} &0  \\
		-\fr{i}{2}e^{i\om \ta +im\si} & \fr{i}{2}e^{-i\om \ta +im\si} &0  &\fr{1}{\sqrt{2}} 
	\end{array}
	\right) \,.\ea
\end{gather}
The corresponding    solution of the reduced theory is\foot{The  
 $\mu$ parameter of the reduced theory here is identified as $\kappa$.}
\be\la{koi}
	g_0=\left(
	\begin{array}{cc}
		g_A& \mbf{0}_4  \\
		\mbf{0}_4& g_S
	\end{array}
	\right) \,, \ \ \ \  \ \ \ \ \    v\equiv  \ept\,,\ee
\bd\ba{c}
	g_A=\left(\ba{cccc}
  i&0&0&0
  \\0&-i&0&0
  \\0&0&i&0
  \\0&0&0&-i	\ea\right)\,,\hs{10pt}
	g_S=\left(\ba{cccc}
	 0 & \fr{\om}{\ka} v & -i\fr{m}{\ka} v  & 0
	\\  -\fr{\om}{\ka} v^*  &0 &0 &  i\fr{m}{\ka} v^*
	\\  i\fr{m}{\ka} v  &0 &0 &  -\fr{\om}{\ka} v 
	\\ 0& -i\fr{m}{\ka} v^* & \fr{\om}{\ka} v^* & 0
	\ea\right)\,,  \ea\ed
\be 	 \ba{c}
	A_{+0}=\left(\ba{ccccc}
	{\bf 0}_{4}&~&{\bf 0}_{4}& ~ &~
	\\ ~& i\left( \fr{m^2}{\ka}-\fr{\ka}{2} \right) & 0 & 0  & 0
	\\ {\bf 0}_{4}& 0 & -i\left( \fr{m^2}{\ka}-\fr{\ka}{2} \right) & 0  & 0
	\\ ~& 0 & 0 & i\left( \fr{m^2}{\ka}-\fr{\ka}{2} \right)  & 0
	\\ ~ &0 & 0 & 0  & -i\left( \fr{m^2}{\ka}-\fr{\ka}{2} \right)
	\ea\right) \,,\\
	A_{-0}=\left(\ba{ccccc}
	{\bf 0}_{4}&~&{\bf 0}_{4}& ~ &~
	\\ ~& -i\fr{\ka}{2} & 0 & 0  & 0
	\\ {\bf 0}_{4}&0 & i\fr{\ka}{2}  & 0  & 0
	\\ ~&  0 & 0 & -i\fr{\ka}{2}   & 0
	\\ ~ &0 & 0 & 0  & i\fr{\ka}{2}
	\ea\right)\,, \ea
\ee
\be 	\YRo=\YLo=0\,. \la{tiki}  \ee
Note that the 
 point-like string (BMN  vacuum) solution is a  particular case
 of  \eqref{rxs3hom}, that is when 
  $m=0$ and $\om=\ka$. 
  In the reduced  theory 
  the  corresponding  limit of \rf{koi} is 
  a  special case of the vacuum in  \eqref{vacuum}.\foot{
One may also consider a  formally different embedding 
of the string solution \eqref{rxs3hom} into the reduced theory 
for which the point-like limit  corresponds to the trivial vacuum 
$g= {\bf 1}$. In this case the solution for $g$  has $\sigma$
instead of $\tau$ dependence, 
see Appendix D.}

Since  the classical fermionic fields vanish, 
 the bosonic $AdS_5$ sector, the bosonic $S^5$ sector and the fermionic sector 
 all decouple at the level of the action and 
  we can discuss them  separately.

Here the $AdS_5$ part of $g_0$ lives in $H$ and is constant.\foot{In the $AdS_5$ 
   case we shall assume  that  the field is  just in the top left $4\x4$ matrix
    of the original $(8\x8)$ field and  similarly for the $S^5$  case 
     the field  will be  just in the bottom right $4 \x4$ matrix.} 
     As discussed in section \ref{vac} this is a vacuum solution of this sector. 
     The resulting fluctuation Lagrangian in the bosonic $AdS_5$ sector is
\be\la{huh}
	L_{A}=\Str \left[ \ha \dpl \h \dm \h  -\de A_-\dpl \h +\de A_+ g_0
	 \dm \h \gi _0 +\de A_+ \de A_- -\gi _0\de A_+g_0\de A_- 
	+\ka ^2\left(  \h\h T ^2 -\h  T  \h T \right)\right] 
\ee
We partially fix the $H$ gauge symmetry by setting the diagonal components of $\h^\perp$ to zero.\foot{This is to completely remove the degeneracy of expanding around this vacuum.} After integrating out $\de A_\pm$ the Lagrangian describing only the physical fluctuations is 
\be\la{lala}
	L_{A}=\Str \left[ \ha \dpl \h^\parallel \dm \h^\parallel
	+\ka ^2\left(  \h^{\parallel}\h^{\parallel} T ^2 -\h^\parallel  T  \h^\parallel T \right)\right] \,.
\ee
Let us introduce the component fields of $\h^\parallel $   as 
\be
	\h^\parallel =\left(
	\begin{array}{cccc}
		0 & 0 & a_{1}+ia_{2}& a_{3}+ia_{4}\\
		0 & 0 & a_{3}-ia_{4}& -a_{1}+ia_{2}\\
		a_{1}-ia_{2}& a_{3}+ia_{4}& 0 & 0 \\
		a_{3}-ia_{4}& -a_{1}-ia_{2}& 0 & 0  
	\end{array}
	\right)\,.
\ee
Then \rf{lala} becomes 
\be 
	L_{A}  =2 \sum_{i=1}^4 \left( \dpl a_i \dm a_i - \ka ^2 a_i^2  \right)\,,
\ee
which describes four bosonic fluctuations with frequency
$
\sqrt{n^2+\ka^2}, \ \  n\in\mbb{Z}.
$

Now let us consider  the $S^5$ sector. 
We introduce the following parametrization of $\h^\parallel$, $\h^\perp$ and $\de A_\pm$,
\be
	\h^\parallel =\left(
	\begin{array}{cccc}
		0 & 0 & b_{1}+ib_{2}& b_{3}+ib_{4}\\
		0 & 0 & -b_{3}+ib_{4}& b_{1}-ib_{2}\\
		-b_{1}+ib_{2}& b_{3}+ib_{4}& 0 & 0 \\
		-b_{3}+ib_{4}& -b_{1}-ib_{2}& 0 & 0  
	\end{array}
	\right) \,,
\ee
\be
\h^\perp=\left(\ba{cccc}
i h_1 & h_2+i h_3&0&0
\\-h_2+i h_3&-i h_1&0&0
\\0&0&i h_4 & h_5+i h_6
\\0&0&-h_5+i h_6&-i h_4\ea\right)\, ,\ee
\be\ba{c}
\de A_+=\left(\ba{cccc}
i a_{+1} & \left(a_{+2}+i a_{+3}\right)v^2&0&0
\\-\left(a_{+2}-i a_{+3}\right)v^{*}{}^2&-i a_{+1}&0&0
\\0&0&i a_{+4} & \left(a_{+5}+i a_{+6}\right)v^2
\\0&0&-\left(a_{+5}-i a_{+6}\right) v^{*}{}^2&-i a_{+4}\ea\right)\, ,
\\
\de A_-=\left(\ba{cccc}
i a_{-1} & a_{-2}+i a_{-3}&0&0
\\-a_{-2}+i a_{-3}&-i a_{-1}&0&0
\\0&0&i a_{-4} & a_{-5}+i a_{-6}
\\0&0&-a_{-5}+i a_{-6}&-i a_{-4}\ea\right)\, .\ea\ee
When we substitute this into the bosonic part of the quadratic
 fluctuation Lagrangian, \rf{quafluc} the fields decouple into two
 smaller sectors. These are, firstly,  a sector  containing $b_3$,
 $b_4$ and the diagonal components of $\h^\perp$, $\de A_\pm$, which has
  a Lagrangian with constant coefficients, and secondly, a sector 
  containing  $b_1$,$b_2$ and the off-diagonal components of 
  $\h^\perp$, $\de A_\pm$. The coefficients in this sector have some
   $\tau$
   dependence, arising from the $\de A_+ \de A_-$ term, ($v$ defined in \rf{koi}
   depends on $\tau$).

If the gauge field fluctuations are integrated out first,  we 
end up with a Lagrangian that has $\tau$-
dependent coefficients. To avoid this complication, i.e. 
to construct an  action containing only physical fluctuations  and  having  
constant coefficients we choose the following partial gauge fixing 
\be\ba{c}
h_1+h_4=\trm{const}\,,
\\\ka (a_{-2}-a_{-5})-\ka^2 (h_3-h_6)-\dm(a_{+3}-a_{+6})-\ka\dm(h_2-h_5)=0 \,,
\\ \ka (a_{-3}+a_{-6})+\ka^2 (h_2+h_5)-\dm(a_{+2}+a_{+5})-\ka\dm(h_3+h_6)=0\,.\ea \la{ghj}
\ee 
Then 
we can easily integrate out the diagonal components of $\de A_\pm$ to get a Lagrangian for $b_3$ and $b_4$ in the desired form.
The second two gauge constraints are chosen to decouple $b_1$ and $b_2$ from the unphysical fluctuations. 
By using the remaining gauge freedom
we should be able to ensure that the unphysical fields only give a trivial contribution  to the partition function.


The resulting Lagrangian for this sector is then
\be 
	\begin{split}
		L_{S}  = & 2 \Big[\sum_{i=1}^4 \dm b_i \dpl b_i + \sum_{i=1}^2 (2m^2- \ka ^2 ) b_i^2 +4 m^2b_4^2 
	+ 2\ka (b_4 \dpl b_3   +b_4 \dm b_3  ) \Big] \,. 
	\end{split}
\ee
This Lagrangian describes two decoupled fluctuations, $b_1$, $b_2$, with frequencies
\be \sqrt{n^2+\ka^2-2m^2}\,,\hs{50pt}n\in\mbb{Z}\,, \la{jjj} \ee 
and two coupled fluctuations, $b_3$, $b_4$,  with frequencies
\be
\sqrt{n^2+2\ka^2-2m^2\pm2\sqrt{n^2\ka^2+(m^2-\ka^2)^2}}\,,\hs{50pt}n\in\mbb{Z}\,. \la{fff}
\ee
In appendix E we shall present an alternative way of computing 
these  fluctuation  frequencies  which does not involve the above gauge fixing, \rf{ghj}.

The  fermionic sector is described by 
\be \label{quafer}\ba{c}
\!\!\!\!\!\!\!\!\!\!\!\!\!\!\!\!\!\!\!\!\!\!\!\!\!\!\!\!\!\!\!\!\!\!\!\!\!\!\!\!\!\!\!\!\!\!\!\!\!\!\!\!\!\!\!\!\!\!\!\!\!\!\!\!\!\!\!\!\!\!\!
	L_{ferm}={\rm STr} \left( \ \ha \de \Psi_{_R} \commut{T}{\dm \de \Psi_{_R}+
	\commut{A_{0-}}{\de \Psi_{_R}}}\right.\\ \left.\hs{100pt}+ \ha \de \Psi_{_L} \commut{T}{\dpl \de \Psi_{_L}+
	\commut{A_{0+}}{\de \Psi_{_L}}} + \ka \gi_0 \de \Psi_{_L} g_0\de \Psi_{_R}   \right) \,.\ea
\ee
To  make  coefficients in this   Lagrangian constant 
we may  rotate some  of the fermionic fields  
to cancels  the contribution of $g_0$ and $\gi_0$
 in the ``Yukawa'' interaction term.
This can be achieved by parametrizing   the  matrix components of  $\de \Psi_{_R}$ and $\de \Psi_{_L}$
as follows
\begin{gather}
	\de \Psi_{_R}=\left(
	\begin{array}{cc}
		0& \mf{X}_R \\
		\mf{Y}_R& 0 
	\end{array}
	\right) \,, ~~~
	\de \Psi_{_L}=\left(
	\begin{array}{cc}
		0& \mf{X}_L  \\
		\mf{Y}_L& 0 
	\end{array}
	\right) \,,
\end{gather}
where
\begin{gather}
	\mf{X}_R=\left(
	\begin{array}{cccc}
		 0 & 0 &\alpha_1+i\alpha_2 &\alpha_3+i\alpha_4 \\
		 0 & 0 &-\alpha_3+i\alpha_4 &\alpha_1-i\alpha_2 \\
		\alpha_5+i\alpha_6 &\alpha_7-i\alpha_8 &0&0\\
		\alpha_7+i\alpha_8 &-\alpha_5+i\alpha_6 &0 &0 \\
	\end{array}
	\right)\,, \\
	\mf{Y}_R=\left(
	\begin{array}{cccc}
		0 & 0 &-\alpha_6 -i\alpha_5 & -\alpha_8 -i\alpha_7 \\
		0 & 0 &\alpha_8 -i\alpha_7 & -\alpha_6 +i\alpha_5 \\
		\alpha_2 +i\alpha_1 & \alpha_4 -i\alpha_3 &0 &0  \\
		\alpha_4 +i\alpha_3 & -\alpha_2 +i\alpha_1&0 &0  
	\end{array}
	\right)\,,
\end{gather} 
\begin{gather}
	\mf{X}_L=\left(
	\begin{array}{cccc}
		0 & 0 &(\beta_1+i\beta_2)v^*  &(\beta_3+i\beta_4)v  \\
		0 & 0 &(\beta_3-i\beta_4 )v^*  &(-\beta_1+i\beta_2)v \\
		(\beta_5+i\beta_6 )v^*  &(-\beta_7+i\beta_8)v &0&0\\
		(\beta_7+i\beta_8 )  v^*  &(\beta_5-i\beta_6)v   &0 &0 
	\end{array}
	\right)\,,  \\
	\mf{Y}_L=\left(
	\begin{array}{cccc}
		0 & 0 &(-\beta_6 -i\beta_5)v &( -\beta_8 -i\beta_7)v \\
		0 & 0 &(-\beta_8 +i\beta_7 )v^*  &( \beta_6 -i\beta_5)v^*  \\
		(\beta_2 +i\beta_1)v & (-\beta_4 +i\beta_3)v &0 &0  \\
		(\beta_4 +i\beta_3)v^* & (\beta_2 -i\beta_1 )v^* &0  &0  
	\end{array}
	\right)\,. 
\end{gather}
Here $\alpha_k$ and $\beta_k$  are 8+8 real anticommuting functions and $v$ is defined in \rf{koi}.
The Lagrangian \eqref{quafer} then takes the form 
\be\ba{l}
	L_{ferm}=2 \Big[ \sum _{i=1}^8\left( \alpha_i \dm \alpha_i +\beta_i \dpl \beta_i \right) \nonumber \\ 
	\hs{20pt}
	+ \sqrt{\ka^2+m^2}\left(-\alpha_1 \alpha_2 + \alpha_3 \alpha_4 -\alpha_5 \alpha_6 -\alpha_7 \alpha_9  
	+\beta_1 \beta_2-\beta_3 \beta_4+\beta_5 \beta_6+\beta_7 \beta_8 \right)\nonumber \\ 
	\hs{20pt}
	+\sqrt{\ka^2-m^2}\left( \alpha_1 \beta_3 + \alpha_3 \beta_1 -\alpha_5 \beta_7 -\alpha_7 \beta_5  
	-\beta_2 \alpha_4+\beta_4 \alpha_2+\beta_6 \alpha_8-\beta_8 \alpha_6 \right)  \Big] \,,
\ea\ee
which describes 8 fermionic fluctuations with 4+4  sets of the frequencies,
\begin{equation}
	\begin{split}
	&\sqrt{n^2-m^2+\fr{5\ka ^2}{4}+\sqrt{\ka^4+n^2\ka^2-m^2\ka^2}}\,,\\
	&\sqrt{n^2-m^2+\fr{5\ka ^2}{4}-\sqrt{\ka^4+n^2\ka^2-m^2\ka^2}}\,,\hs{50pt}n\in\mbb{Z}\,.
	\end{split}
\end{equation}
The  characteristic frequencies found above directly from the reduced theory action 
 are exactly the same as found \ci{fpt,art}
 from the \adss string theory action   expanded near the solution \rf{rxs3hom}.\foot{Starting
  with the string solution 
in the form \rf{rxs3hom} used in \ci{art} one finds that the fermions are naturally periodic
 \ci{btz}.}
 

We conclude that expanding  the superstring action 
near  the homogeneous 2-spin solution in $\mbb{R}_t\x S^3$
and expanding the reduced theory action near its counterpart in the reduced theory one finds the same set of 
characteristic frequencies and thus the same  one-loop contribution
to the respective partition functions.

 \subsubsection{Large spin limit of the  folded spinning string in $AdS_3 \times S^1$} 
 
 As another example we shall  consider  the  large spin limit of 
 the solution  for a folded  string in $AdS_5$  with spin $S$ \ci{gkp}  orbiting also 
 in $S^5$   with momentum $J$ \ci{ft1}. 
 As was noticed in \ci{ftt,rtt}, in the limit when ${\cal S }= \fr{S}{ \sqrt \lambda} \to \infty $
 with $\fr{J}{\sqrt \lambda\ \ln S}$ 
 fixed \ this solution simplifies  and becomes  homogeneous. 
In terms of the embedding coordinates (see appendix \ref{Coord})
it takes the form (cf. \rf{rxs3hom})
\be\la{gkp}
\ba{c}
Y_0+iY_{-1}=\cosh (\mm\sigma) \,e^{i\ka \ta}\,, ~~~~~~ Y_1+iY_2= \sinh (\mm\sigma) \,e^{i\ka \ta}\,, ~~~~~~  Y_3=Y_4=0\,,
\\X_1=X_{2}= X_3=X_{4}=0\,,~~~~~~X_5+iX_6=e^{i\nu \tau}\,, \ \ \ \   \ka^2=\mm^2+\nu^2  \ , 
\ea\ee
where it is assumed that $\ka \sim \ell \gg 1$, and $\fr{\nu}{\ka}$ is fixed 
(so that the closed-string periodicity condition in $\sigma$ is satisfied asymptotically).
 This solution is, in fact,  related to
the $J_1=J_2$  solution in $\mbb{R}_t \times S^3$ by a formal analytic continuation \ci{rtt}.


Using the parametrization in terms of the embedding coordinates discussed in appendix B
we obtain the corresponding coset element $f$, 
\begin{gather}
\ba{c}
	f=\left(
	\begin{array}{cc}
		f_A& \mbf{0}_4  \\
		\mbf{0}_4& f_S
	\end{array}
	\right) \,,
\\	f_A=\left(
	\begin{array}{cccc}
	e^{\frac{i\ka \tau}{2}}\cosh \frac{\mm\sigma}{2} & 0 &0 &-e^{\frac{3i\ka \tau}{2}}\sinh \frac{\mm\sigma}{2} \\
		0 &e^{\frac{i\ka \tau}{2}}\cosh \frac{\mm\sigma}{2} &e^{-\frac{i\ka \tau}{2}}\sinh \frac{\mm\sigma}{2}  &0 \\
		0 & e^{\frac{i\ka \tau}{2}}\sinh \frac{\mm\sigma}{2} &e^{-\frac{i\ka \tau}{2}}\cosh \frac{\mm\sigma}{2} &0  \\
		-e^{-\frac{3i\ka \tau}{2}}\sinh \frac{\mm\sigma}{2} & 0 &0  &e^{-\frac{i\ka \tau}{2}}\cosh \frac{\mm\sigma}{2}
	\end{array}
	\right) \,,
\\	f_S=\left(
	\begin{array}{cccc}
		e^{\fr{i\nu \ta}{2}} & 0 &0 & 0 \\
		0 & e^{\fr{i\nu \ta}{2}} &0 & 0 \\
		0 & 0 &e^{-\fr{i\nu\ta}{2}}&0\\
		0 & 0 &0 &e^{-\fr{i\nu \ta}{2}} 
	\end{array}
	\right) \,.\ea
\end{gather}
The counterpart of this    solution  in  the reduced theory is  described   
by\foot{ 
Here the $\mu$ parameter of the reduced theory is identified as $\nu$.}
\be\la{koi2}
	g_0=\left(
	\begin{array}{cc}
		g_A& \mbf{0}_4  \\
		\mbf{0}_4& g_S
	\end{array}
	\right) \,, \ \ \ \  \ \ \ \ \    v\equiv  e^{-i\frac{\ka^2\tau}{\nu}} \,,\ee
\bd\ba{c}

g_A=\left(\ba{cccc}
	 0 & \fr{\ka}{\nu} v & -\fr{\mm}{\nu} v  & 0
	\\  -\fr{\ka}{\nu} v^*  &0 &0 &  \fr{\mm}{\nu} v^*
	\\  \fr{\mm}{\nu} v  &0 &0 &  -\fr{\ka}{\nu} v 
	\\ 0& -\fr{\mm}{\nu} v^* & \fr{\ka}{\nu} v^* & 0
	\ea\right) \,,\hs{10pt}
	\ \ \ g_S=\left(\ba{cccc}
  i&0&0&0
  \\0&-i&0&0
  \\0&0&i&0
  \\0&0&0&-i	\ea\right)\,,\ea\ed

\be 	
 \ba{c}
	A_{+0}=\left(\ba{ccccc}
	\frac{i\nu}{2}&0&0&0&~
\\	0&-\frac{i\nu}{2}&0&0&{\bf 0}_{4}
\\0&0&\frac{i\nu}{2}&0&~
\\0&0&0&-\frac{i\nu}{2}&~
\\ ~&{\bf 0}_{4}&~&~&{\bf 0}_{4}\ea\right)\,,
\\
	A_{-0}=\left(\ba{ccccc}
	\frac{i(m^2+\ka^2)}{2\nu}&0&0&0&~
\\	0&-\frac{i(\mm^2+\ka^2)}{2\nu}&0&0&{\bf 0}_{4}
\\0&0&\frac{i(\mm^2+\ka^2)}{2\nu}&0&~
\\0&0&0&-\frac{i(\mm^2+\ka^2)}{2\nu}&~
\\ ~&{\bf 0}_{4}&~&~&{\bf 0}_{4}\ea\right)\,,\ea \la{koii}
\ee
\be 	\YRo=\YLo=0\,. \ee

Note that again the 
 point-like string (BMN  vacuum) solution is a  particular case
of  \eqref{gkp}, that is when 
  $\mm=0$ and $\nu=\ka$. 
  The  corresponding  limit of \rf{koi2} is related by a simple $H$ gauge transformation to 
  a  special case of the vacuum in  \eqref{vacuum}.
 
 This reduced theory background  is very similar to the one in \rf{koi} corresponding 
 to the homogeneous string solution in $\mbb{R}_t \x S^3$. 
 Carrying out a similar analysis of the quadratic fluctuation spectrum in the reduced theory action 
 one finds the 
  following bosonic 
 \be\ba{c}
 1 \hs{10pt} \x \hs{10pt} \sqrt{n^2+2\ka^2+ 2\sqrt{\ka^4+n^2\nu^2}}\,,
 \\1 \hs{10pt} \x \hs{10pt} \sqrt{n^2+2\ka^2-2\sqrt{\ka^4+n^2\nu^2}}\,,
 \\2 \hs{10pt} \x \hs{10pt} \sqrt{n^2+2\ka^2-\nu^2}\,,
 \\4 \hs{10pt} \x \hs{10pt} \sqrt{n^2+\nu^2}  \,
\ea \ee
and  fermionic 
\be\ba{c}
4 \hs{10pt} \x \hs{10pt}\sqrt{n^2+\ka^2+\frac{\nu^2}{4}+\sqrt{\nu^2(n^2+\ka^2)}}\,,
\\4 \hs{10pt} \x \hs{10pt} \sqrt{n^2+\ka^2+\frac{\nu^2}{4}-\sqrt{\nu^2(n^2+\ka^2)}}
\,
\ea\ee
fluctuation frequencies. 
These are indeed exactly the same  as following directly from the  \adss superstring  
  action, \ci{ftt}.


\renewcommand{\theequation}{5.\arabic{equation}}
 \setcounter{equation}{0}
 
\section{Concluding  remarks\la{Comm}}

In this  paper we  discussed how to relate the semiclassical 
expansion in the  original  \adss superstring  theory \rf{fulllag}  and the corresponding  
reduced theory \rf{Ltot}. We  considered several 
classes of string solutions,  found their reduced model counterparts 
and then  verified that   the respective spectra of quadratic fluctuations
match. This implies  the matching of the one-loop partition functions \rf{eq}.

Given that the classical equations (and their solutions) 
in  the string theory   and in the reduced theory are closely related, 
one may, of course,    expect  the quadratic fluctuations to match as well. However,
this matching is still  rather  non-trivial given that one  needs to 
partially fix the $H \times H$ gauge symmetry of the string equations
written in terms of the reduced theory
variables \rf{redeom} in order to be able to construct  a local Lagrangian 
of  the reduced theory.  One of the remaining  open questions 
is if the reduced Lagrangians 
obtained via different gauge fixings, (in particular, the ones parametrized by an
automorphism $\tau$, see \rf{redgaugefields},\rf{gauwzw}),   are actually equivalent 
at the quantum level. 

 It would be interesting  to understand the  equivalence 
 between the corresponding quadratic fluctuation spectra in the  string theory and 
 in the reduced theory using their closely  connected integrable structures. 
 Indeed,   fluctuation frequencies near  particular  finite gap solutions can be found 
 directly from the corresponding algebraic curve description (see, e.g., \ci{grom}).

 Another  important open problem is to find out  if  the one-loop  matching between the 
 string and the reduced  theory partition functions  extends  to the two-loop level. 
 If it does,  that  would be a truly   non-trivial  confirmation of  our conjecture \rf{eq}.
 On the string theory side,  the two-loop computation of the partition function was  done 
 for the infinite spin (or ``homogeneous'') limit of the folded string solution \ci{rtt,rt2}. 
 What remains  is to compute the two-loop  correction  starting with the reduced theory action \rf{Ltot}
 and expanding it near the corresponding solution \rf{koi2},\rf{koii}. 
 Since the analysis   of  quadratic  fluctuations on the reduced  theory side is generally simpler 
 than on the string theory side 
 we expect  that this two-loop computation may
 not be too complicated, (cf. also \ci{rtfin}).

Finally, as a step towards a solution of the  reduced theory  based on its integrability 
it remains  to compute the  corresponding 2-d   Lorentz-invariant massive S-matrix 
for the  elementary excitations above the ``trivial''  vacuum. 
There are technical complications  when this is done  directly by starting with 
the reduced theory based on the symmetrically
gauged ($\tau = {\bf 1}$) WZW model \rf{gauwzw} expanded   near the 
vacuum $g = {\bf 1}$. However,
 one may try to expand  near other  vacua  like \rf{vacuum}
or consider a reduced  model with a non-trivial automorphism $\tau$
(expecting still that the S-matrix should not depend on 
a choice of $ \tau  $).

\section*{Acknowledgements}
We acknowledge useful discussions  with A. Alexandrov, M. Grigoriev, 
  R.  Roiban  and A. Tirziu.  
 AAT is particularly grateful to R.  Roiban for 
 an  initial  collaboration on
  the issues addressed in this paper and numerous discussions of related questions. 
BH would like to thank EPSRC for his studentship. 
YI is supported by the Ishizaka Foundation.
Part of this work was done while   AAT was a participant of the 2009 
program ``Fundamental Aspects of Superstring Theory'' 
at the  Kavli Institute for Theoretical Physics at Santa Barbara.  
AAT  also acknowledges the hospitality of the Galileo Galilei Institute in Florence 
during the  2009 program  "Non-Perturbative Methods in Strongly Coupled Gauge Theories".


\appendix

\renewcommand{\theequation}{A.\arabic{equation}}
 \setcounter{equation}{0}

\section{
 $\Psu$: some definitions and notation \la{PSU}}

Here we will present a  particular matrix representation  of  $\Psu$ which we used 
in the main text  (see also \ci{review,gt1}).
 In particular, we shall make explicit  the identification of the $\mf{g}=\mf{sp}\left(2,
\,2\right)\x\mf{sp}\left(4\right)$ subalgebra  whose 
corresponding group $G$ is the subgroup $G$ in the $F/G$ coset sigma model, 
and also the group $G$ in the $G/H$ gauged WZW model.

Let us  define the following matrices
\be
\mbf{\Si}=\left(\ba{cc}\Si&\mbf{0}_4\\\mbf{0}_4&\mbf{1}_4\ea\right)\,,
\ \ \ 
\mbf{K}=\left(\ba{cc}K&\mbf{0}_4\\\mbf{0}_4&K\ea\right)\,, \ \ \ \ \ \Si^2=\mbf{1}_4, \ \ \ K^2=-\mbf{1}_4\,,
\ee
\bd
\Si=\left(\ba{cccc}1&0&0&0\\0&1&0&0\\0&0&-1&0\\0&0&0&-1\ea\right)\,,
\hs{30pt}
K=\left(\ba{cccc}0&-1&0&0\\1&0&0&0\\0&0&0&-1\\0&0&1&0\ea\right)\,.
\ed
 We can then write a generic element of the algebra $\psu$ as follows
\be \mf{f}=\left(\ba{cc}\mf{A}&\mf{X}\\\mf{Y}&\mf{B}\ea\right)\,,\ee
\bd \mf{f}=-\mbf{\Si}^{-1}\mf{f}^{\dagger}\mbf{\Sigma}\,,\hs{30pt}\trm{Tr }\mf{A}=\trm{Tr }\mf{B}=0\,,
\ed
\bd \mf{f}^{\dagger}=\left(\ba{cc}\mf{A}^{\dagger}&-i\mf{Y}^{\dagger}\\-i\mf{X}^{\dagger}&\mf{B}^{\dagger}\ea\right)\,.\ed
Here $\mf{A}$ and $\mf{B}$ are $4\x4$ matrices whose components are commuting  while
 $\mf{X}$ and $\mf{Y}$ are $4\x4$ matrices whose components are anticommuting. We then have
  the following conditions on $\mf{A},\;\mf{B},\;\mf{X}$ and $\mf{Y}$,
\be
\Si\mf{A}^{\dagger}\Si=-\mf{A}\,,\hs{30pt}
\mf{B}^{\dagger}=-\mf{B}\,,\hs{30pt}
i\Si \mf{Y}^{\dagger}=\mf{X}\,,\hs{30pt}
i\mf{X}^{\dagger}\Si=\mf{Y}\,.
\ee
Thus  $\mf{A}\in\mf{su}\left(2,\,2\right)$ and 
 $\mf{B}\in\mf{su}\left(4\right)$. 
We can then decompose $\mf{f}$ under a $\mbb{Z}_4$ grading as follows
\be \mf{f}=\mf{f}_0\oplus\mf{f}_1\oplus\mf{f}_2\oplus\mf{f}_3\,,\ee
\bd-\mbf{K}^{-1}\mf{f}_r^{st}\mbf{K}=i^r \mf{f}_r\,,\hs{40pt}\mf{f}_r^{st}=\left(\ba{cc}\mf{A}^t&-\mf{Y}^t\\\mf{X}^t&\mf{B}\ea\right)\,.\ed
It is possible to write generic elements of $\mf{f}_{0,2}$ as
\be \mf{f}_{0,2}=\left(\ba{cc}\mf{A}_{0,2}&\mbf{0}_4\\\mbf{0}_4&\mf{B}_{0,2}\ea\right)\,,\ee
\bd\ba{c}
\mf{A}_0=K\mf{A}_0^t K\,,\hs{30pt}\mf{B}_0=K\mf{B}_0^t K\,,
\\\mf{A}_2=-K\mf{A}_2^t K\,,\hs{30pt}\mf{B}_2=-K\mf{B}_2^t K\,,
\ea\ed
and generic elements of $\mf{f}_{1,3}$ as
\be \mf{f}_{1,3}=\left(\ba{cc}\mbf{0}_4&\mf{X}_{1,3}\\\mf{Y}_{1,3}&\mbf{0}_4\ea\right)\,,\ee
\bd
i\mf{X}_1=-K\mf{Y}_1^t K\,,\hs{30pt}i\mf{X}_3=K\mf{Y}_3^t K\,.
\ed
The subspaces of this decomposition satisfy the following commutation relations
\be\la{com} \left[\mf{f}_i,\,\mf{f}_j\right]\subset\mf{f}_{i+j \trm{ mod } 4}\,. \ee
We identify $\mf{f}_0=\mf{g}$ and $\mf{f}_2=\mf{p}$.
Then  $\mf{g}$ forms a subalgebra, and it is this algebra whose 
corresponding group is the group $G$ in the
 $F/G$ coset sigma model and in the $G/H$ gauged WZW model.

 It is now possible to perform a further $\mbb{Z}_2$ decomposition, which allows us to define the group $H$ in the $G/H$ gauged WZW model. To do this we identify the following fixed element $T\in \mf{f}_2$
\be
T=\fr{i}{2}\trm{ diag}\left(1,\, 1,\, -1,\, -1,\, 1,\, 1,\, -1,\, -1\right)\,.\ee
The $\mbb{Z}_2$ decomposition is then given by
\be\mf{f}^\parallel_r=-\left[T,\left[T,\mf{f}_r\right]\right]\,,\hs{40pt} \mf{f}^\perp_r=-\{T,\{T,\mf{f}_r\}\}\,.\ee
It should be noted that this is an orthogonal decomposition, that is
\be\ba{c}
\mf{f}=\mf{f}^\parallel\oplus\mf{f}^\perp\,,\\
\trm{STr}(\mf{f}^\parallel \mf{f}^\perp)=0\,.\ea
\ee
Then 
\be
\left[\mf{f}^\perp,\,\mf{f}^\perp\right]\subset\mf{f}^\perp\,,\hs{30pt}
\left[\mf{f}^\perp,\,\mf{f}^\parallel\right]\subset\mf{f}^\parallel\,,\hs{30pt}
\left[\mf{f}^\parallel,\,\mf{f}^\parallel\right]\subset\mf{f}^\perp\,.\ee
We identify $\mf{h}=\mf{f}_0^\perp$, $\mf{m}=\mf{f}_0^\parallel$, $\mf{a}=\mf{f}_2^\perp$, 
$\mf{n}=\mf{f}_2^\parallel$. Elements from these subspaces  satisfy 
\be
\left[\mf{a},\mf{a}\right]\subset 0\,, \hs{20pt}
\left[\mf{a},\mf{h}\right]\subset 0\,, \hs{20pt}
\left[\mf{h},\mf{h}\right]\subset \mf{h}\,, \hs{20pt}
\left[\mf{m},\mf{m}\right]\subset \mf{h}\,, \hs{20pt}
\left[\mf{m},\mf{h}\right]\subset \mf{m}\,, \hs{20pt}
\left[\mf{m},\mf{a}\right]\subset \mf{n}\,, \hs{20pt}
\left[\mf{n},\mf{a}\right]\subset \mf{m} \ee
Here $\mf{h}$ is a subalgebra; 
the corresponding 
subgroup is then identified as the group $H$ in
 the $G/H$ gauged WZW model.
  It is possible to  show that $\mf{h}$ has the following form
\be\left(\ba{cccc}
\mf{h}_1&\tbf{0}_2&\tbf{0}_2&\tbf{0}_2
\\\tbf{0}_2&\mf{h}_2&\tbf{0}_2&\tbf{0}_2
\\\tbf{0}_2&\tbf{0}_2&\mf{h}_3&\tbf{0}_2
\\\tbf{0}_2&\tbf{0}_2&\tbf{0}_2&\mf{h}_4\ea\right)\,,\ee
where each $\mf{h}_i$ is a copy of $\mf{su}\left(2\right)$,
i.e.
 $\mf{h}=\mf{su}\left(2\right)\oplus\mf{su}\left(2\right)\oplus\mf{su}\left(2\right)
 \oplus\mf{su}\left(2\right)\cong\mf{so}\left(4\right)\oplus\mf{so}\left(4\right)$.

 Finally as discussed in \ci{review}, it is possible to use the
  \kasym to choose fermionic currents to take  the form,
\be\la{kapsym}\left(\ba{cccccccc}
0&0&0&0&0&0&\bullet&\bullet
\\0&0&0&0&0&0&\bullet&\bullet
\\0&0&0&0&\bullet&\bullet&0&0
\\0&0&0&0&\bullet&\bullet&0&0
\\0&0&\bullet&\bullet&0&0&0&0
\\0&0&\bullet&\bullet&0&0&0&0
\\\bullet&\bullet&0&0&0&0&0&0
\\\bullet&\bullet&0&0&0&0&0&0
\ea\right)\,.\ee
This is exactly the same as the structure of the
 fermionic elements of the $^\parallel$ space. 
 Thus it is always possible to choose the
  \kasym gauge such
   that the fermionic currents live in the $^\parallel$ space.

\renewcommand{\theequation}{B.\arabic{equation}}
 \setcounter{equation}{0}

\section{ Parametrization in terms of  embedding coordinates 
\la{Coord}}

Here we  shall discuss the relation between the embedding coordinates in 
$AdS_5 \x S^5$  and parametrization of the corresponding  $\Psu$ coset elements
(see \cite{review} for details).
 
Let us  define six real coordinates $Y^{M}$ on $\mbb{R}^{4,2}$
\ ($M=-1,\;0,\ldots,\;4$)  and six real coordinates $X^{I}$ on $\mbb{R}^6$\  ($I=1,\;2,\;\ldots,\;6$).
 To define $AdS_5$ and $S^5$ embedded in $\mbb{R}^{4,2}$ and $\mbb{R}^6$  we impose
\be\la{norm}
\ba{cc}
\h_{MN}^{4,2}Y^M Y^N=-1\,,&\h_{IJ}^{6,0}X^I X^J=1\,,
\\\h^{4,2}=\trm{diag}\left(-1,\;-1,\;1,\;1,\;1,\;1\right)\,,& \ \ \ 
\h^{6,0}=\trm{diag}\left(1,\;1,\;1,\;1,\;1,\;1\right)\,.
\ea
\ee
Finally we define another set of coordinates, $t,\;y_i$ on $AdS_5$ and $\thet,\;x_i$ on $S^5$, $i=1,\;2,\;3,\;4$:
\be
Y^1+i Y^2=\fr{y_1+i y_2}{1-\fr{y^2}{4}}\,, \hs{40pt} Y^3+i Y^4=\fr{y_3+iy_4}{1-\fr{y^2}{4}}\,,
\ee
\bd
Y^0+iY^{-1}=\fr{1+\fr{y^2}{4}}{1-\fr{y^2}{4}}e^{it}\,,
\ed
\be
X^1+i X^2=\fr{x_1+ix_2}{1+\fr{x^2}{4}}\,,\hs{40pt}X^3+i X^4=\fr{x_3+ix_4}{1+\fr{x^2}{4}}\,,
\ee
\bd
X^5+i X^6=\fr{1-\fr{x^2}{4}}{1+\fr{x^2}{4}}e^{i\thet}\,.
\ed
Here  $y^2=y_i y_i$ and $x^2=x_i x_i$.
The corresponding metrics of $AdS_5$ and $S^5$ in terms of $t,\;y_i,\;\thet,\;x_i$ are 
\be\ba{c}\la{met}
\h_{MN}^{4,2}dY^M  dY^N =-\left(\fr{1+\fr{y^2}{4}}{1-\fr{y^2}{4}}\right)^2 dt^2+\fr{dy_i dy_i}{\left(1-\fr{y^2}{4}\right)^2}\,,
\\\h_{IJ}^{6,0}dX^I dX^J=\left(\fr{1-\fr{x^2}{4}}{1+\fr{x^2}{4}}\right)^2 d\thet^2+\fr{dx_i dx_i}{\left({1+\fr{x^2}{4}}\right)^2}\,.
\ea
\ee
A suitable choice of bosonic coset element would be 
 such that  $\Str\left(f^{-1}df\right)^2$ 
 coincides with the sum of the two metrics in \rf{met}. 
 This  allows us to relate
 the embedding  coordinates 
 with the bosonic coset element directly: 
\be\la{embedding}
\ba{c}
f=\left(\ba{cc}f_A&\mbf{0}_4\\\mbf{0}_4&f_S\ea\right)
\\=\left(\ba{cc}\trm{exp}\left(\fr{i}{2}t\ga_5\right)&\mbf{0}_4\\\mbf{0}_4&\trm{exp}\left(\fr{i}{2}\thet\ga_5\right)\ea\right)
\left(\ba{cc}\fr{1}{\sr{1-\fr{y^2}{4}}}\left(\mbf{1}_4+\fr{1}{2}y_i\ga_i\right)&\mbf{0}_4\\
\mbf{0}_4&\fr{1}{\sr{1+\fr{x^2}{4}}}\left(\mbf{1}_4+\fr{i}{2}x_i\ga_i\right)\ea\right)\ea\,.
\ee
Here $\ga_k$ are   the  $\mf{so}\left(5\right)$ Dirac matrices  chosen as 
\be
\ga_1=\left(\ba{cccc}
0&0&0&-1
\\0&0&1&0
\\0&1&0&0
\\-1&0&0&0\ea\right)\,,\hs{20pt}
\ga_2=\left(\ba{cccc}
0&0&0&i
\\0&0&i&0
\\0&-i&0&0
\\-i&0&0&0\ea\right)\,,\hs{20pt}
\ga_3=\left(\ba{cccc}
0&0&1&0
\\0&0&0&1
\\1&0&0&0
\\0&1&0&0\ea\right)\,,
\ee
\bd
\ga_4=\left(\ba{cccc}
0&0&-i&0
\\0&0&0&i
\\i&0&0&0
\\0&-i&0&0\ea\right)\,,\hs{20pt}
\ga_5=\left(\ba{cccc}
1&0&0&0
\\0&1&0&0
\\0&0&-1&0
\\0&0&0&-1\ea\right)\,.
\ed

\subsection{$AdS_2 \x S^2$\la{coord2x2}}

Let us now consider a special case  of an 
$AdS_2 \x S^2$ subspace of  $AdS_5\x S^5$:
\be\la{norm2x2}\ba{c}
-\left(Y^{-1}\right)^2-\left(Y^0\right)^2+\left(Y^1\right)^2=-1\,,
\\\left(X^1\right)^2+\left(X^5\right)^2+\left(X^6\right)^2=1\,,
\\Y_2=Y_3=Y_4=y_1=y_3=y_4=0\,,
\\X_2=X_3=X_4=x_1=x_3=x_4=0\,.
\ea\ee
The explicit 
 coordinates on $AdS_2 \x S^2$
  are $t,\;y=y_1,\;\thet,\;x=x_1$.

The corresponding 
parametrization of the  $PSU(2,2|4)$ element, \rf{embedding}, 
is then 
\be\ba{c}
f_{A}=\sr{1-\fr{y^2}{4}}\left(\ba{cccc}
e^{\fr{it}{2}}&0&0&\fr{iy}{2}e^{\fr{it}{2}}
\\0&e^{\fr{it}{2}}&\fr{iy}{2}e^{\fr{it}{2}}&0
\\0&-\fr{iy}{2}e^{-\fr{it}{2}}&e^{-\fr{it}{2}}&0
\\-\fr{iy}{2}e^{-\fr{it}{2}}&0&0&e^{-\fr{it}{2}}
\ea\right)\,,
\\f_{S}=\sr{1+\fr{x^2}{4}}\left(\ba{cccc}
e^{\fr{i\thet}{2}}&0&0&-\fr{x}{2}e^{\fr{i\thet}{2}}
\\0&e^{\fr{i\thet}{2}}&-\fr{x}{2}e^{\fr{i\thet}{2}}&0
\\0&\fr{x}{2}e^{-\fr{i\thet}{2}}&e^{-\fr{i\thet}{2}}&0
\\\fr{x}{2}e^{-\fr{i\thet}{2}}&0&0&e^{-\fr{i\thet}{2}}\ea\right)\,.
\ea\ee
Following the prescription of Pohlmeyer reduction as
 discussed in section $\ref{Class}$, we 
 can 
 make a $G$ gauge transformation, $f_{b}\ra f_{b} g'$,
  such that $\left(f_{b}^{-1}\dpl f_{b}\right)_\mf{p}\in\mf{a}$.
   We can then use the remaining conformal diffeomorphism invariance 
    to set $\left(f_{b}^{-1}\dpl f_{b}\right)_\mf{p}=\mu_+ T$. In terms of the embedding 
    coordinates this then implies
\be\ba{c}
-\left(\dpl Y^{-1}\right)^2-\left(\dpl Y^0\right)^2+\left(\dpl Y^1\right)^2=-\mu_+^2\,,
\\\left(\dpl X^1\right)^2+\left(\dpl X^5\right)^2+\left(\dpl X^6\right)^2=\mu_+^2\,.
\ea\ee
The next step  of the  reduction is to find a element $g_0$ of $G$
 such that $\left(f_{b}^{-1}\dm f_{b}\right)_\mf{p}$ $=\mu_- \gi_0 T g_0$. 
 The following element of $G$ satisfies this relation
\be
g_0=\left(\ba{cc}
g_A&\mbf{0}_4
\\\mbf{0}_4&g_S\ea\right)\,,
\ee
\bd
g_A=\left(\ba{cccc}
i\cosh\phi_A&0&0&\sinh\phi_A
\\0&-i\cosh\phi_A&\sinh\phi_A&0
\\0&\sinh\phi_A&i\cosh\phi_A&0
\\\sinh\phi_A&0&0&-i\cosh\phi_A
\ea\right)\,,
\ed\bd
g_S=\left(\ba{cccc}
i\cos\phi_S&0&0&i\sin\phi_S
\\0&-i\cos\phi_S&i\sin\phi_S&0
\\0&i\sin\phi_S&i\cos\phi_S&0
\\i\sin\phi_S&0&0&-i\cos\phi_S
\ea\right)\,,\ed
provided  
the following relations are satisfied 
\be\ba{c}
-\left(\dm Y^{-1}\right)^2-\left(\dm Y^0\right)^2+\left(\dm Y^1\right)^2=-\mu_-^2\,,
\\\left(\dm X^1\right)^2+\left(\dm X^5\right)^2+\left(\dm X^6\right)^2=\mu_-^2\,,
\ea\ee
\be\ba{c}
-\dpl Y^{-1}\dm Y^{-1}-\dpl Y^{0}\dm Y^{0}+\dpl Y^1 \dm Y^1=-\mu^2\cosh 2\phi_A\,,
\\\dpl X^1 \dm X^1+\dpl X^5 \dm X^5+\dpl X^6 \dm X^6=\mu^2 \cos 2\phi_S\,,
\\\mu^2=\sr{\mu_+^2\mu_-^2}\,.
\ea\ee
It is possible to check  that the corresponding gauge fields $A_\pm$
in  \rf{redgaugefields}  vanish in this case.


\renewcommand{\theequation}{C.\arabic{equation}}
 \setcounter{equation}{0}

\section{ Fluctuations near $AdS_2 \x S^2$ solutions: \   special cases\la{Check}}

In section \ref{2x2} it was shown that for a classical solution  in $AdS_2 \x S^2$
 the bosonic fluctuation equations are
\be\ba{c}\la{fluct1}
\dpm \ze_i+\mu^2\cosh2\phi_A\;\ze_i=0\,,\hs{40pt}i=1,\,\ldots,\,4\,,
\\\dpm \ze_i+\mu^2\cos 2\phi_S\;\ze_i=0\,,\hs{40pt}i=5,\,\ldots,\,8\,,
\ea\ee
and the fermionic fluctuation equations are given by the following sets of coupled equations
\be
\ba{cc}\la{fluct3}
\dm \vt_i +\mu\,\cos\phi_S\cosh\phi_A\,{\vt'}_i+\mu\,\sin\phi_S\sinh\phi_A\,{\vt'}_{i+1}=0\,,&
\\\dpl {\vt'}_i-\mu\,\cos\phi_S\cosh\phi_A\,\vt_i+\mu\,\sin\phi_S\sinh\phi_A\,\vt_{i+1}=0\,,&\hs{30pt}i=1,\,3,\,5,\,7
\\\dm \vt_{i+1}+\mu\,\cos\phi_S\cosh\phi_A\,{\vt'}_{i+1}-\mu\,\sin\phi_S\sinh\phi_A\,{\vt'}_i=0\,,&
\\\dpl {\vt'}_{i+1}-\mu\,\cos\phi_S\cosh\phi_A\,\vt_{i+1}-\mu\,\sin\phi_S\sinh\phi_A\,\vt_i=0\,.&
\ea
\ee
Below we shall consider some special cases of these equations.

\subsection{Giant Magnon}

Here we shall check the general claim that the above equations give 
the same one-loop correction as  the calculation  following directly from the string theory  action written in terms of coordinates on \adss with the  example 
   of  the giant magnon
  solution \ci{hm,giantmagsc}. 
For the giant magnon string solution 
we decompactify the spatial worldsheet direction
(the energy and angular 
   momentum of the string are taken to infinity).
Its counterpart in the reduced 
theory 
is  the vacuum and kink solutions of the  sinh-Gordon and sine-Gordon equations respectively 
\be\ba{c}
\phi_A=0\,,
\\\phi_S=2\arctan e^{\fr{\sigma-v\tau}{\sr{1-v^2}}}\,.
\ea\ee 
When taking the large energy/spin limit 
we rescale the worldsheet coordinates by $\mu$ and then send  $\mu\ra\infty$. 
As a result,  $\mu$ scales out of the fluctuation equations 
\rf{fluct1}-\rf{fluct3}. We may also change to  the Lorentz-boosted coordinates
\be
\Si=\fr{\si-v\ta}{\sr{1-v^2}}\ , \hs{30pt}\T=\fr{\si-v\ta}{\sr{1-v^2}} \ . 
\ee
 The bosonic $AdS_5$ fluctuation equations are given by 4 copies of 
\be\la{adsgiantmag}
\dpm\ze_A+\ze_A=0\,.
\ee
The bosonic $S$ fluctuation equations are given by four copies of
\be\la{sgiantmag}
\dpm\ze_S+\left(1-2\sech^2\Si\right)\ze_S=0\,.
\ee
As discussed in \ci{giantmagsc} to compute the one-loop fluctuation operator  determinant 
for the giant magnon string solution
we should first look for the plane-wave solutions
 of the fluctuation equations. The plane-wave solutions of \rf{adsgiantmag} are proportional to
\be\la{adsplwa}
\ze_A=e^{i k\Si+i \om \T}\,,\hs{50pt}\om^2=k^2+1\,, 
\ee
and of \rf{sgiantmag} to 
\be\la{splwa}
\ze_S =e^{i k\Si+i \om \T}\left(\tanh\Si+i k\right)\,,\hs{50pt}\om^2=k^2+1\,.
\ee
Finally, the fermionic fluctuation equations are given by eight copies of
\be \ba{c}
\dm\vt-\tanh\Si\,\tvt=0\,,
\\\dpl\tvt+\tanh\Si\,\vt=0\,.
\ea\ee
After  some simple manipulation with expressions in 
 \ci{giantmagsc} it is easy to see that this system has plane-wave solutions proportional to
\be\la{plwaferm}
\ba{c}
\vt=-\fr{\left(1-v\right)\left|\om-k\right|}{\sr{1-v^2}}e^{ik\Si-i\om\T}e^{\fr{i}{2}\left(\arctan\left(-\om\sinh2\Si\right)-\arctan\left(k\tanh2\Si\right)\right)}\sech\Si\sr{\left|\om\cosh2\Si-k\right|}\,,
\\\tvt=e^{ik\Si-i\om\T}e^{\fr{i}{2}\left(\arctan\left(\om\sinh2\Si\right)-\arctan\left(k\tanh2\Si\right)\right)}\sech\Si\sr{\left|\om\cosh2\Si+k\right|}\,,
\\\om^2=k^2+1\,.
\ea
\ee
Following $\ci{giantmagsc}$ we may then compute 
the stability angles for these solutions. To do this we put the system in a box of length $L\gg1$, with $\si\sim\si+L$. From the form of the classical solution the system is also periodic in time with period $T_p=\fr{L}{v}$. The stability angle $\nu$ of an arbitrary fluctuation $\de\phi$ is defined to be
\be
\de\phi\left(\ta+T_p,\si\right)=e^{-i\nu}\de\phi\left(\ta,\si\right)\,.
\ee
From \rf{adsplwa} the four stability angles from the bosonic $AdS_5$ sector are
\be
\nu_k\left(\zeta_A\right)=\fr{L}{v}\fr{\om+vk}{\sr{1-v^2}}\,.
\ee
From \rf{splwa} the four stability angles from the bosonic $S^5$ sector are
\be
\nu_k\left(\zeta_S\right)=\fr{L}{v}\fr{\om+vk}{\sr{1-v^2}}+2\cot^{-1}k\,.
\ee 
Finally, from \rf{plwaferm} the eight stability angles from the fermionic sector are
\be
\nu_k\left(\vt,\,\tvt\right)=\fr{L}{v}\fr{\om+vk}{\sr{1-v^2}}+\cot^{-1}k\,.
\ee 
These agree exactly with the results in $\ci{giantmagsc}$
derived directly by 
considering  fluctuations of  coordinates on  $AdS_5 \times S^5$. 
 We then reproduce the 
  final result of $\ci{giantmagsc}$
  that  the sum over the stability angles (with a negative sign for the fermionic contribution)
   vanishes and thus so does the one-loop correction to the 
   logarithm of the partition function or  the energy of the giant magnon.

\subsection{Some other 
 examples\la{kink}}

Here we briefly consider some other interesting solutions in $AdS_2 \x S^2$.
As discussed in \ci{hosv} the reduced theory solutions 
\be\la{sol1}
\phi_S=\trm{am}\left(\fr{\mu\left(\ta-v \si\right)}{k\sr{1-v^2}},k^2\right)\,,\hs{40pt}
\phi_A=0\,,
\ee
and
\be\la{sol2}
\phi_S=\fr{\pi}{2}+\trm{am}\left(\fr{\mu\left(\si-v \ta\right)}{k\sr{1-v^2}},k^2\right)\,,\hs{40pt}
\phi_A=0\,,
\ee
give rise to single-spin helical strings, effectively living on $\mbb{R}_t \x S^2$.\foot{$\trm{am}$ is the Jacobi amplitude function.}

These solutions include some special cases. For example if we take 
the $v\ra 0$ limit in \rf{sol1} the corresponding string solution is a
 string pulsating on $S^2$, which is also discussed in \ci{kt}. If we take the
  $k\ra \infty$, $\mu \ra\infty$, $\frac{\mu}{k}\ra 1$ limit of \rf{sol2} we get 
  the sine-Gordon kink solution, which, as previously discussed, corresponds 
  to the giant magnon string solution \ci{hm}.

For both \rf{sol1} and \rf{sol2} the bosonic fluctuation equations from the $AdS_5$ sector are trivial, as we just have the vacuum solution. For the $S^5$ sector we obtain four copies of the following equations:
\be\la{flucs1}\left(\dpm +\mu^2\left[2\trm{cn}^2\left(\fr{\mu\left(\ta-v
\si\right)}{k\sr{1-v^2}},k^2\right)-1\right]\right)\zeta_S=0 \ee
for \rf{sol1} and 
\be\la{flucs2}\left(\dpm +\mu^2\left[1-2\trm{cn}^2 \left(\fr{\mu\left(\si-v \ta\right)}{k\sr{1-v^2}},k^2
\right) \right]\right)\zeta_S=0 \ .\ee
for \rf{sol2}.
These are strongly related to the $n=1$ Lam$\acute{\trm{e}}$ equation, \ci{ww}. For the fermionic sector
 we obtain eight copies of the following coupled systems
\be\ba{c}\la{flucferm1}\dm \vt  +\mu\  \trm{cn}\left(\fr{\mu\left(\ta-v \si\right)}{k\sr{1-v^2}},k^2
\right)\ \tvt=0
\\\dpl \tvt -\mu\ \trm{cn}\left(\fr{\mu\left(\ta-v \si\right)}{k\sr{1-v^2}},k^2\right)\ \vt=0 \ea\ee
for \rf{sol1} and 
\be\ba{c}\la{flucferm2}\dm \vt  -\mu\  \trm{sn}\left(\fr{\mu\left(\si-v \ta\right)}{k\sr{1-v^2}}
,k^2\right)\ \tvt=0
\\\dpl \tvt +\mu\ \trm{sn}\left(\fr{\mu\left(\si-v \ta\right)}{k\sr{1-v^2}},k^2\right)\ \vt=0\ea\ . \ee
for \rf{sol2}.
In various special cases the spectra and 
determinants of these operators have been studied 
in much detail, \ci{ww,paw,db}. Therefore,
 it should be possible to  compute  the corresponding  one-loop 
 correction  to the logarithm of the partition function at least numerically. 

\renewcommand{\theequation}{D.\arabic{equation}}
 \setcounter{equation}{0}

\section{ Examples of reduced theory counterparts of some simple string  solutions}

Here  we shall consider the  reduced theory counterparts of the homogeneous string 
solutions on $\mbb{R}_t \times S^3$ and $AdS_3 \times S^1$. Compared to the  discussion 
 in section 4.3 we shall assume the trivial embedding of these solutions 
 into the reduced theory when  it can be truncated to the complex  sine-Gordon
 or complex sinh-Gordon models respectively. 
The bosonic part of the reduced theory  counterpart of 
 $AdS_3 \times S^3$  string theory is described by, ($\del_\pm = \del_\tau \pm \del_\s$), \foot{This  Lagrangian is found by starting with the reduced
 theory based on 
 the symmetrically  gauged $G/H = SO(1,2)/SO(2) \times SO(3)/SO(2) $ gWZW model
 and integrating out the $SO(2) \times SO(2)$ gauge fields  \ci{gt1}.}
\bea
  && L_B =
   \del_+ \varphi \del_- \varphi  + 
    \cot^2 { \varphi }\ \del_+ \theta \del_- \theta \no \\ 
 &&\ \ \ \   +\ 
   \del_+ \phi \del_- \phi  +   \coth^2 { \phi }\ \del_+ \chi  \del_- \chi
    +\frac{\m^2}{2} (\cos 2\varphi  - \cosh 2\phi) .   \la{syy}
\eea
A particular simple solution  of the resulting equations of motion is \foot{ More
 general solutions  of CSG were discussed in \ci{okamura,hosv}. }
\bea
&&    \vp =\vp_0, \ \ \  \ \ \ \phi=\phi_0, \ \ \ \ \ \
\theta= n \s + a \tau, \ \ \ \ \ \   \chi=  k \s + b \tau \ , \la{sop}\\
&&\,\,\,\,\,\,\,\,\, \m^2 \sin^4 \vp_0  = n^2 -a^2 \ , \ \ \ \ \ 
 \ \ \   \m^2 \sinh^4 \p_0  = k^2 - b^2 \ . 
\eea 
In the case of 
the $J_1=J_2$ homogeneous string  solution in $\mbb{R}_t \times S^3$  \rf{rxs3hom}
we have 
\be \la{str}
t= \k \tau,\ \ \ \ \ \
X_1=   \frac{ 1}{\sqrt 2}    e^{ i ( w  \tau + m \s)} ,\ \ \ \ \
  X_2= \frac{ 1}{ \sqrt 2}   e^{ i ( w  \tau - m \s)} \ , \ \ \ \ \ \
\k^2 \equiv \mu^2 =  w^2 + m^2 \ .   \ee
In the  reduced theory we have 
 $\mu^2 \cos 2 \vp  =  \del_+  X_i  \del_- X^*_i $,
 so that the corresponding solution  has 
   $\varphi=\vp_0 = \const.$,  with 
\be \la{gh}
\cos 2 \vp_0 = \frac{w^2 - m^2}{ w^2 + m^2}\ , \ \ \ \ \ \ \ \ \
\sin \vp_0 =  \fr{m}{\mu} \ , \ \ \ \ \cos \vp_0 =  \fr{w}{\mu}   \ . 
\ee
Also,  for $
\theta  = n \s  + a \tau $, 
the equation  of motion for $\phi$  implies 
\be \la{uy} \mu^2 \sin^4 \vp_0 = n^2- a^2  \ ,  \ \ {\rm i.e.} \ \ 
\fr{ m^4}{ w^2 + m^2 } =  n^2- a^2   \ . 
\ee
Note that here we cannot set $n=0$. If $\sigma$ is periodic $n$ should be an integer, which imposes constraints on $m$ and $w$. 
A special   solution with $w=0$  (i.e. $J=0$)   
corresponds to  $\vp_0 = \fr{\pi}{ 2} $ and $m=n$. 

The  embedding of the  circular string solution 
into the reduced model  considered in \rf{koi} 
was different -- it contained only 2-d time dependence. 
Note that  had we started with the axially gauged  $SO(3)/SO(2) $  WZW 
model the $\cot^2 { \varphi }$ in the kinetic term would be replaced  
by $\tan^2 { \varphi }$ and the $a^2$ and $n^2$   terms in  \rf{uy}
would change places. In this case we could get a solution of the reduced theory which looks more like that found in \rf{koi}. 

Indeed, if we replace $v$ by $e^{i\theta}$, $\fr{\om}{\ka}$ by $\cos \varphi$ and
 $\fr{m}{\ka}$ by $\sin \varphi$ in $\rf{koi}$ and then integrate out $A_\pm$ at 
 a classical level, we get  the complex sine-Gordon model with $\tan^2 { \varphi }$ 
 in the kinetic term.  This is also related to the fact that the point-like
or BMN  limit of the above solution ($m \to 0$) corresponds to 
the trivial vacuum in the reduced theory (see \cite{gt2}),
 which was not the case in \rf{koi}.\foot{ 
It should  be noted that here the gauge group, $SO(2)$, is abelian and thus the axial gauging,
 $\tau(u)=-u$, is allowed. For non-abelian groups this is not possible as such a 
 map  $\tau$ is no 
 longer an automorphism of the algebra. Instead,  we may  use automorphisms like those discussed
  in section \ref{vac}.
}


\

For the   homogeneous  solution in   $AdS_3 \times S^1$,  \rf{gkp},
(i.e.  the limit of  large $\k$ and large $\ell$  when we can ignore periodicity of $\s$), 
we have   $  \del_+ Y_0  \del_- Y_0  + \del_+ Y_{-1}  \del_- Y_{-1}
    -  \del_+ Y_1 \del_- Y_1-  \del_+ Y_2 \del_- Y_2
      =      {\m^2} \cosh 2\phi $  \ ($\mu=\nu$).
 Then
 \be \la{hjl}
 \k^2 + \ell^2 =  {\m^2} \cosh 2\phi_0 \ , \ \ \ \ \sinh \p_0  =   \fr{ \ell}{ \mu} \ , \ \ \ \ 
 \cosh \phi_0 =  \fr{ \k}{ \mu }  \ , \ \ \ \  \m= \sqrt{ \k^2 - \ell^2}   \ .   
 \ee    
Thus  the  solution is 
\bea  
 \p = \p_0\ , \ \ \ \  \ \ \ \ \chi = k  \s  + b \tau \ , \ \ \ \ \ \ \ \ \ \ \
  k^2 - b^2 = \mu^2 \sinh^4 \p_0 \ , \ \ \  \ \ \ \ \ \sinh \p_0  =   \fr{ \ell}{ \mu} \ .
 \eea
In the scaling limit  $k$  and $m$  need not be integers. 
As long as  we decompactify $\s$ we can always rotate $b$ to 0 by 
a 2-d Lorentz transformation   since this is a symmetry of the reduced theory. Also,
$k$ needs to be non-zero.

Starting with the axially gauged or  ``T-dual'' model with 
$\coth^2 \p \to \tanh^2 \p$  would interchange $k$ and $b$. 
Again, 
the  reduced theory  embedding of the solution \rf{gkp} discussed in \rf{koi2},\rf{koii}
was  only $\tau$-dependent   and thus  was different.

\renewcommand{\theequation}{E.\arabic{equation}}
 \setcounter{equation}{0}

\section{ An alternative  computation of  reduced theory  fluctuation frequencies
 }

In section 4.3 we partially fixed the $H$ gauge symmetry such that two of the physical 
fluctuation fields in $\eta^\parallel$ decoupled from the remaining 
unphysical fluctuation fields.
However,  this strategy may  not necessarily work for other homogeneous solutions, e.g.,
 the  ``small'' spinning string  in $\mbb{R}_t \x S^5$ discussed in \ci{ft3,ft2}. 
Here  we shall use the example of the $S^5$ sector of the reduced theory solution 
corresponding to the two-spin homogeneous string in $\mbb{R}_t \x S^3$, (section 4.3.1),
 to discuss an alternative strategy for computing the characteristic frequencies.

We introduce the following parametrization of $\h^\parallel$, $\h^\perp$ and $\de A_\pm$,
\be
	\h^\parallel =\left(
	\begin{array}{cccc}
		0 & 0 & b_{1}+ib_{2}& b_{3}+ib_{4}\\
		0 & 0 & -b_{3}+ib_{4}& b_{1}-ib_{2}\\
		-b_{1}+ib_{2}& b_{3}+ib_{4}& 0 & 0 \\
		-b_{3}+ib_{4}& -b_{1}-ib_{2}& 0 & 0  
	\end{array}
	\right) \,,
\ee
\be
\h^\perp=\left(\ba{cccc}
i h_1 & h_2+i h_3&0&0
\\-h_2+i h_3&-i h_1&0&0
\\0&0&i h_4 & h_5+i h_6
\\0&0&-h_5+i h_6&-i h_4\ea\right)\, ,\ee
\be\ba{c}
\de A_+=\left(\ba{cccc}
i a_{+1} & 0&0&0
\\0&-i a_{+1}&0&0
\\0&0&i a_{+4} & 0
\\0&0&0&-i a_{+4}\ea\right)\, ,
\\
\de A_-=\left(\ba{cccc}
i a_{-1} & a_{-2}+i a_{-3}&0&0
\\-a_{-2}+i a_{-3}&-i a_{-1}&0&0
\\0&0&i a_{-4} & a_{-5}+i a_{-6}
\\0&0&-a_{-5}+i a_{-6}&-i a_{-4}\ea\right)\, .\ea\ee
Using the $H$ gauge freedom we set
 the off-diagonal components    of $\de A_+$ to zero. 
When we substitute 
these expressions   into the bosonic part of the quadratic fluctuation Lagrangian \rf{quafluc}, 
we get a Lagrangian with constant coefficients.
 As in section 4.3.1
the fields decouple into two smaller sectors,  the  first  sector containing $b_3$, $b_4$ and 
the diagonal components of $\h^\perp$, $\de A_\pm$, and  the second  sector 
containing  $b_1$,$b_2$ and the off-diagonal components of $\h^\perp$, $\de A_\pm$.
  
We can easily integrate out the diagonal components of $\de A_\pm$, and then 
end up with a Lagrangian for 
$14$ fields ($4$ of $\h^\parallel$, $6$ of $\h^\perp$, $4$ of $\de A_-$), 
some of which are unphysical. 
Using the fact that we have the two decoupled sectors, we can split the corresponding $14 \times 
14$ mass matrix into two parts, a
$4 \times 4$ matrix containing $b_{3}$ and $b_{4}$, and a $10 \times 10$ matrix containing $b_1$ and $b_2$.

Substituting $e^{i(\Omega \tau- n\sigma)}$ into the equations of motion we find that the $4 
\times 4$ matrix takes the form,
\be
\left(
\begin{array}{cccc}
 4 \left(n^2-\Omega^2\right) & 8 i \kappa \Omega & -\frac{2 m \left(n^2-\Omega^2\right)}{\sqrt{
 \kappa^2-m^2}} & -\frac{2 m \left(n^2-\Omega^2\right)}{\sqrt{\kappa^2-m^2}} \\
 -8 i \kappa \Omega & -4 \left(4 m^2-n^2+\Omega^2\right) & \frac{4 i \kappa m \Omega}{\sqrt{
 \kappa^2-m^2}} & \frac{4 i \kappa m \Omega}{\sqrt{\kappa^2-m^2}} \\
 \frac{2 m \sqrt{\kappa^2-m^2} \left(n^2-\Omega^2\right)}{-\kappa^2+m^2} & -\frac{4 i \kappa m
  \Omega}{\sqrt{\kappa^2-m^2}} & -\frac{m^2 \left(n^2-\Omega^2\right)}{-\kappa^2+m^2} & -\frac{m^2
   \left(n^2-\Omega^2\right)}{-\kappa^2+m^2} \\
 \frac{2 m \sqrt{\kappa^2-m^2} \left(n^2-\Omega^2\right)}{-\kappa^2+m^2} & -\frac{4 i \kappa m 
 \Omega}{\sqrt{\kappa^2-m^2}} & -\frac{m^2 \left(n^2-\Omega^2\right)}{-\kappa^2+m^2} & -\frac{m^2
  \left(n^2-\Omega^2\right)}{-\kappa^2+m^2}
\end{array}
\right) \,.
\ee
This matrix has rank $2$, i.e. it has two non-vanishing eigenvalues. 
The resulting  two  charactersitic frequencies are then found to be the same 
as those in \rf{fff} in 
section 4.3.1,
\be 
\sqrt{n^2+2\ka^2-2m^2\pm2\sqrt{n^2\ka^2+(m^2-\ka^2)^2}}\,,\hs{50pt}n\in\mbb{Z}\,. \la{kol}
\ee 
The $10 \times 10$ matrix has rank $10$. The condition that its  determinant vanishes gives the
 following characteristic fluctuation frequencies,  
\be\label{phys3}
2\  \x  \ \ \sqrt{n^2+\ka^2-2m^2}\,,\hs{50pt}n\in\mbb{Z}\,,
\ee 
and
\be
\begin{split}\label{unphys3}
2 \ \times \  \ n\pm \kappa  \,,\hs{100pt} \\
2 \ \times \ \ n\pm \frac{\kappa^2-2 m^2}{\kappa}  \,.\hs{50pt}n\in\mbb{Z}\,.
\end{split}  
\ee
The frequencies in \eqref{phys3} are the same as those in \rf{jjj} in section 4.3.1, (i.e. 
like  \rf{kol} they match the 
frequencies found from the conformal-gauge string theory). 

The frequencies in \eqref{unphys3} 
give a trivial ($\kappa,\,m$-independent) 
contribution  to the one-loop partition function that  should be cancelled against 
ghost (or path integral measure) terms. 


The approach employed here, i.e.  evaluating a larger mass matrix including unphysical 
fluctuations in addition to physical fluctuations, 
should also be applicable to other homogeneous solutions.  
In particular, we can  apply  it  to  the homogeneous string solution discussed
in section 4.3.2. 
However, it is not clear  whether it may be useful for extending the computation 
to the two-loop level as the unphysical modes, (which we did not explicitly 
decouple above, as this was irrelevant at the one-loop level),  may get coupled through the
interaction terms. 


\end{document}